\documentclass[referee]{jfm}
\usepackage{graphicx}
\usepackage{natbib}

\ifCUPmtlplainloaded \else
  \checkfont{eurm10}
  \iffontfound
    \IfFileExists{upmath.sty}
      {\typeout{^^JFound AMS Euler Roman fonts on the system,
                   using the 'upmath' package.^^J}%
       \usepackage{upmath}}
      {\typeout{^^JFound AMS Euler Roman fonts on the system, but you
                   dont seem to have the}%
       \typeout{'upmath' package installed. JFM.cls can take advantage
                 of these fonts,^^Jif you use 'upmath' package.^^J}%
       \providecommand\upi{\pi}%
      }
  \else
    \providecommand\upi{\pi}%
  \fi
\fi

\ifCUPmtlplainloaded \else
  \checkfont{msam10}
  \iffontfound
    \IfFileExists{amssymb.sty}
      {\typeout{^^JFound AMS Symbol fonts on the system, using the
                'amssymb' package.^^J}%
       \usepackage{amssymb}%
         \let\leq=\leqslant
         
      }{}
  \fi
\fi

\ifCUPmtlplainloaded \else
  \IfFileExists{amsbsy.sty}
    {\typeout{^^JFound the 'amsbsy' package on the system, using it.^^J}%
     \usepackage{amsbsy}}
    {}
\fi

\newsavebox{\astrutbox}
\sbox{\astrutbox}{\rule[-5pt]{0pt}{20pt}}

\title[Heterogeneous vapour condensation in boundary layers]{Theory of surface deposition 
from boundary layers containing condensable vapour and particles}

\author[J. C. Neu, A. Carpio, and L. L. Bonilla]%
{J.\ns C.\ns N\ls E\ls U$^1$,\ns
A.\ns C\ls A\ls R\ls P\ls I\ls O$^2$,\ns \break
\and L.\ns L.\ns B\ls O\ls N\ls I\ls L\ls L\ls A$^3$}

\affiliation{$^1$Department of Mathematics, University of California at Berkeley,
Berkeley, CA 94720, USA\\[\affilskip]
$^2$Departamento de Matem\'atica Aplicada, Universidad
Complutense de Madrid, E-28040 Madrid, Spain\\[\affilskip]
$^3$G.\ Mill\'an Institute of Fluid Dynamics, Nanoscience and Industrial 
Mathematics, Universidad Carlos III de Madrid, 28911 Legan\'es, Spain}

\pubyear{2008}
\volume{538}
\pagerange{119--126}
\date{?? and in revised form ??}

\begin{document}

\maketitle

\begin{abstract}
Heterogeneous condensation of vapours mixed with a carrier gas in the stagnation point 
boundary layer flow near a cold wall is considered in the presence of solid particles much 
larger than the mean free path of vapour particles. The supersaturated vapour condenses on 
the particles by diffusion and particles and droplets are thermophoretically attracted to the 
wall. Assuming that the heat of vaporization is much larger than $k_{B}\tilde{T}_{
\infty}$, where $\tilde{T}_{\infty}$ is the temperature far from the wall, vapour 
condensation occurs in a {\em condensation layer} (CL). The CL width and characteristics 
depend on the parameters of the problem, and a parameter $R$ yielding the rate of vapour 
scavenging by solid particles is particularly important. Assuming that the CL is so narrow that 
temperature, particle density and velocity do not change appreciably inside it, an asymptotic 
theory is found, the $\delta$-CL theory, that approximates very well the vapour and droplet 
profiles, the dew point shift and the deposition rates at the wall for wide ranges of the wall
temperature $\tilde{T}_{w}$ and the scavenging parameter $R$. This theory breaks down
for $\tilde{T}_{w}$ very close to the maximum temperature yielding non-zero droplet 
deposition rate, $\tilde{T}_{w,M}$. If the width of the CL is assumed to be zero (0-CL 
theory), the vapour density reaches local equilibrium with the condensate immediately after it 
enters the dew surface. The 0-CL theory yields appropriate profiles and deposition rates in 
the limit as $R\to\infty$ and also for any $R$, provided $\tilde{T}_{w}$ is very close 
to $\tilde{T}_{w,M}$. Nonlinear multiple scales also improve the 0-CL theory, providing 
good uniform approximations to the deposition rates and the profiles for large $R$ or for 
moderate $R$ and $\tilde{T}_{w}$ very close to $\tilde{T}_{w,M}$, but it breaks down 
for other values of $\tilde{T}_{w}$ and small $R$.
\end{abstract}

\section{Introduction}
\label{sec:intro}
The effects of condensation in fluid flows have been studied theoretically and experimentally
in many situations of interest including Prandtl-Meyer flows describing the trailing edge of 
the blades in steam turbines, \cite{del98}, Ludwieg shock-tube experiments, \cite{luo07},
and condensation trail formation in aircraft wakes, \cite{pao04}. During many deposition 
processes, heterogeneous condensation of vapours on particles and transport towards cold walls 
occur. Examples include vapour deposition from combustion gases, 
\cite{CR88,CR89}, fouling and corrosion in biofuel plants, \cite{pyy03}, 
Outside vapour Deposition (OVD) processes used for making optical fibers, 
\cite{fil03,tandon}, chemical vapour deposition, vapour condensation and aerosol 
capture by cold plates or rejection by hot ones, \cite{rosner}. In these situations, 
deposition of particles and condensed vapour in cold walls is
enhanced by thermophoresis which drives particles and droplets towards the plate, 
\cite{BS85,GR86}.

In this paper, we consider heterogeneous condensation of vapours mixed with a carrier gas in 
the stagnation point boundary layer flow near a cold wall. This problem was already studied 
theoretically by \cite{GR86}, \cite{CR88,CR89} and \cite{fil03} in the case of diluted
vapours in a carrier gas and a diluted suspension of solid particles over which the vapour may 
condense. Rosner and coauthors consider the example of Na$_{2}$SO$_{4}$ vapours in air
whereas \cite{fil03} considers deposition of germanium vapours in a mixture of products 
of stoichiometric methane combustion. \cite{CR88,CR89} study a simple thermophysical 
model in which the carrier gas is considered to be incompressible, the Soret and Dufour 
effects are ignored and the particles and droplets move towards the wall by thermophoresis, 
\cite{zheng,davis}. \cite{GR86} and \cite{fil03} deal with more complicated
thermophysical models in which the carrier gas is compressible, its viscosity has an algebraic
dependence with temperature and the Soret effect is included. In all cases, the presence of
vapours and suspended solid particles does not affect the laminar boundary layer flow of the 
carrier gas, which is described by coupled ordinary differential equations in a similarity 
variable. If the heat of vaporization is much larger than the thermal energy (temperature 
times the Boltzmann constant) far from the wall, vapour condensation occurs in a condensation 
layer (CL) whose distance to the wall, width and characteristics depends on the parameters of 
the problem. Outside the CL, the vapour is undersaturated and it cannot condense on the 
solid particles suspended in the carrier gas. In contrast to this dry region, there is a condensation
region closer to the cold wall where condensation on the particles may occur. 

There are different theories describing the condensation region. The simplest theory, due to 
\cite{CR89}, assumes that the with of the CL is zero and that the vapour is in equilibrium 
with the condensed liquid at the dew surface. We call this approximation the 0-CL theory 
and it has the advantage that no specific mechanism of the condensation of supersaturated 
vapour on suspended solid particles needs to be considered. The 0-CL theory is a good 
approximation to the numerical solution of the complete thermophysical model, 
\cite{CR88}, if vapour is scavenged by the solid particles at a high rate before it can 
condense directly on the cold wall. A correction to the 0-CL theory is given by \cite{fil03} 
for his more complete thermophysical model. He uses a method of multiple scales with a 
linear relation between fast and slow spatial variables to describe the CL. This approximation 
is then matched to the numerical solution of the flow problem in the dry region outside the 
CL. The resulting values of the deposition rates are studied for different versions of his 
thermophysical model and compared to similar results obtained with the simplified model of 
incompressible carrier gas. \cite{fil03} does not solve numerically the equations of the 
complete thermophysical model inside the CL and therefore does not compare the results of 
his theory with a numerical solution of the complete model. In addition, \cite{fil03}'s 
multiple scales theory is not consistent for it would lead to contradictory results had the next
 order in this method been carried out.

In this paper, we revisit the 0-CL theory obtaining new formulas for the dew point shift and
the deposition rates and we give two new asymptotic theories, one based on matched
asymptotic expansions (the $\delta$-CL theory) and another on nonlinear multiple scales
(NLMS). These two new theories give a complete description of vapour density profiles and
deposition rates for a wide range of wall temperatures and compare very well with the
numerical solution of the complete model, much better in fact that the 0-CL theory.

To present our theories, we have adopted the simple thermophysical model of \cite{CR88} 
with one change. \cite{CR88} assume that supersaturated vapour condenses on solid 
particles according to the free
molecular regime law. They consider as an example Na$_{2}$SO$_{4}$ vapours in air with a 
diluted suspension of solid particles with radius one micron. In this case, the mean free path 
of vapour particles is three tenths of the particle radius, so we have assumed that the 
supersaturated vapour condenses on the particles by diffusion. We present three different 
asymptotic theories of the condensation process, calculate the shift in the dew point interface 
due to the flow, the vapour density profile and the deposition rates at the wall and compare 
them to direct numerical simulation of the equations governing the model. Firstly, we revisit 
the 0-CL theory in \cite{CR89} and give approximate formulas for the dew point shift, the
location of the dew point interface and the deposition rates. We also find a formula for the 
maximum value of the wall
temperature for which the deposition rate of condensate carried by droplets to the wall is not
zero. This maximum wall temperature is smaller than the dew point temperature in the 
absence of flow. The second theory is based on matched asymptotic expansions and it is a 
good correction to the 0-CL theory at any scavenging rate except when the wall temperature 
is very close to its maximum value for condensate deposition via droplets. Instead of assuming 
that the CL has zero width, we consider a CL of finite width $\delta l_{b}$ (small compared 
with the width of the Hiemenz boundary layer, $l_{b}$, \cite{Hiemenz}) detached from the 
wall, and within which the vapour density has not yet reached local equilibrium with the 
liquid. In this $\delta$-CL theory, the temperature, the flow and thermophoretic velocities, 
and the particle density are constant within the thin CL. This assumption is questionable if
the temperature at the wall, $\tilde{T}_{w}$, is so high that the CL is attached to the wall, 
which occurs for  $\tilde{T}_{w}$ near the maximum temperature yielding non-zero 
droplet deposition rate, $\tilde{T}_{w,M}$. In fact, for $\tilde{T}_{w}$ slightly
below  $\tilde{T}_{w,M}$, the $\delta$-CL theory yields unrealistic values of the 
deposition rates. The third theory is a nonlinear multiple scales (NLMS) method which 
corrects the 0-CL theory for high vapour scavenging rates and is free from the inconsistencies of 
\cite{fil03}'s theory. We have compared the results of the three asymptotic theories (in 
particular the approximate deposition rates at the wall they provide) to a numerical solution 
of the complete model. In the limit as $l^2_{b}$ times the radius of the solid particles is 
large compared to the reciprocal of the particle number density (large vapour scavenging by 
particles), the method of nonlinear multiple scales and the $\delta$-CL theory yield very 
good approximations to the vapour density and deposition fluxes at the wall. For moderate 
and low scavenging rates, the $\delta$-CL theory provides the best approximations to the 
vapour and droplet density profiles and deposition rates at the wall except for 
$\tilde{T}_{w}$ slightly below $\tilde{T}_{w,M}$. For such values of $\tilde{T}_{w}$,
the NLMS theory is the best approximation.

Even though we have presented asymptotic results for the thermophysical model by
\cite{CR88} (which is relatively simple and can be numerically solved at a lower 
computational cost), our theories should also apply to the more complete and computationally
costlier thermophysical models by \cite{GR86} and \cite{fil03} for which the comparisons
with numerical results are much more expensive.
 
The rest of the paper is as follows. Section \ref{sec:model} describes the thermophysical
model we use. In Section \ref{sec:stagnation}, the equations of the model are particularized 
to the simple case of stagnation point flow. We also derive exact expressions for the deposition
rates. Section \ref{sec:equilibrium} contains the results of using the 0-CL theory which 
considers the vapour to be in equilibrium with the condensed liquid at the dew surface. If the 
CL is well detached from the wall and its width is small but not zero, we use a description by
means of matched asymptotic expansions in Section \ref{sec:condlayer}: the $\delta$-CL
theory. Section \ref{sec:attached} describes a multiple scales method that is useful when 
there is strong scavenging of vapour by the solid particles and therefore the length needed for 
the vapour concentration to decay to its equilibrium value is very small. Section 
\ref{sec:numerical} contains comparison of the results of our different approximations to
a direct numerical solution of the governing equations for stagnation flow. Lastly Section 
\ref{sec:final} contains a discussion of our results and conclusions. 

\section{Model}
\label{sec:model}
Consider a dilute vapour of number density $\tilde{c}(\mathbf{\tilde{x}})$ in a carrier 
gas that contains a small amount of solid single-size particles. The mass fraction of vapour 
and of solid particles are sufficiently small with respect to the mass fraction of the carrier gas,
so that the velocity and temperature fields (assumed to be {\em stationary}), $\mathbf{
\tilde{u}}(\mathbf{\tilde{x}})$ and $\tilde{T}(\mathbf{\tilde{x}})$, are not 
affected by the condensation and deposition processes. 
The solid particles can act as condensation sites for the vapour. Let $n_{*}$ be the volume of 
a particle divided by the molecular volume of condensed vapour, so that a solid particle is equivalent 
to $n_{*}$ molecules of vapour. Then a droplet of liquid coating a solid particle is equivalent to 
$\tilde{n}(\mathbf{\tilde{x}})$ vapour molecules, in the sense that $\tilde{n}$ equals 
the volume of a droplet 
(particle plus condensed vapour) divided by the molecular volume of condensed vapour. Thus the 
number of liquid molecules coating a given solid particle is $\tilde{n}(\mathbf{\tilde{x}})
-n_{*}$. Let $\tilde{\rho}(\mathbf{\tilde{x}})$ be the number density of droplets, so that 
$\tilde{\rho}(\mathbf{\tilde{x}})\, [\tilde{n}(\mathbf{\tilde{x}})-n_{*}]$ is the 
number density of the condensate. Since the number of droplets equals the number of solid 
particles, the continuity equation for $\tilde{\rho}$ is
\begin{eqnarray}
\tilde{\nabla}\cdot\left[\left(\mathbf{\tilde{u}} - \alpha\nu{\tilde{\nabla}
\tilde{T}\over\tilde{T}}\right)\tilde{\rho}\right] = 0.   \label{1}
\end{eqnarray}
In this equation, the velocity of droplets equals the flow velocity plus the thermophoretic 
velocity which is $-\alpha\nu\tilde{\nabla}\ln\tilde{T}$ ($\nu$ is the kinematic 
viscosity of the carrier gas and $\alpha$ is a dimensionless thermophoretic coefficient which 
depends on the particle radius). We shall assume that the carrier gas is incompressible. This
leads to simpler equations and asymptotic expressions but it also overestimates the particle 
deposition rates, cf. Fig.\ 7 in \cite{fil03}. For wall temperatures larger than $\tilde{
T}_{\infty}/2$, this effect is not too large. Our
asymptotic theories also apply to more realistic models including compressibility of the 
carrier gas. For an incompressible carrier gas, Eq.\ (\ref{1}) yields
\begin{eqnarray}
\left(\mathbf{\tilde{u}} - \alpha\nu{\tilde{\nabla}\tilde{T}\over\tilde{T}}
\right)\cdot\tilde{\nabla}\tilde{\rho} = \alpha\nu\tilde{\rho}\,\tilde{\nabla}
\cdot{\tilde{\nabla}\tilde{T}\over\tilde{T}}.   \label{2}
\end{eqnarray}

The mean free path $\lambda_{vg}$ of vapours diluted in a carrier gas is small compared 
to the size of the particles suspended in the gas. In fact, for Na$_{2}$SO$_{4}$ 
vapours in air, the ratio of their molecular weights is $z=142/28$, so that the mean free path 
$\lambda_{vg}$ of vapours relative that of pure air, $\lambda_{g}$, is, \cite{davis}
\begin{eqnarray}
\lambda_{vg} = \sqrt{{2\over 1+z}}\, {4\, \lambda_{g}\over\left(1+{
\sigma_{v}\over\sigma_{g}}\right)^2}, \label{3}
\end{eqnarray}
where $\sigma_{v}$ and $\sigma_{g}$ are the collision diameters of the vapour and 
of air molecules, respectively. We estimate $\sigma_{g}= 3.7\times 10^{-8}$ cm
(based on the collision diameter of nitrogen) and $\sigma_{v}=5.5\times 10^{-8}$ cm 
(based on the molecular volume of Na$_{2}$SO$_{4}$ in the solid phase). Hence 
$\lambda_{vg}/\lambda_{g}=0.371$, according to Eq.\ (\ref{3}). At $\tilde{T}=300$ K, 
$\lambda_{g}=0.065\,\mu$m, and at $\tilde{T}=1400$ K it is $1400/300$ times this, or 0.3 
$\mu$m. Eq.\ (\ref{3}) yields $\lambda_{vg}= 0.11\,\mu$m. Instead of (\ref{3}), we 
may use the average length over which a vapour molecule randomizes its momentum 
(``loses its sense of direction''), see Eq.\ (8) in \cite{peeters}
\begin{eqnarray}
\lambda_{vg} = \sqrt{\frac{1+z}{2}}\, {4\, \lambda_{g}\over\left(1+{
\sigma_{v}\over\sigma_{g}}\right)^2}, \label{3bis}
\end{eqnarray}
which yields $\lambda_{vg}= 0.34\,\mu$m at $\tilde{T}=1400$ K. This is still relatively
small. Thus we can consider that supersaturated vapour condenses on a spherical particle of 
radius 1 $\mu$m by diffusion. The diffusive flux of vapour diluted in the incompressible 
carrier gas is $\tilde{J}_{v}= D4\upi\tilde{r}^2\partial\tilde{c}/\partial
\tilde{ r}$, which yields $\tilde{c}(\tilde{r})= \tilde{c} -\tilde{J}_{v}/(4\upi D 
\tilde{r})$ provided the flux is constant and $\tilde{c}$ is the vapour density far from the 
droplet whose radius is $a$. At the droplet, $\tilde{c}(a)=\bar{c}<\tilde{c}$, so that the 
diffusive flux towards the droplet is $\tilde{J}_{v}= 4\upi D a (\tilde{c}-\bar{c})$, 
and it should equal the rate at which the droplet captures vapour molecules, $d\tilde{n}/d
\tilde{t}$. In the stationary gas flow we consider, $d\tilde{n}/d\tilde{t}= (\mathbf{
\tilde{u}}-\nu\alpha\tilde{\nabla}\ln\tilde{T})\cdot\tilde{\nabla}\tilde{n}$. 
The simplest model for the vapour concentration at the surface of a 
droplet is that absorption and desorption of vapour molecules is so fast that $\bar{c}=
\tilde{c}_{e}$, the equilibrium number density of vapour. Since 
$a=[3v\tilde{n}/(4\upi)]^{1/3}$ ($v$ is the molecular volume of vapour), we have 
\begin{eqnarray}
\left(\mathbf{\tilde{u}} - \alpha\nu{\tilde{\nabla}\tilde{T}\over\tilde{T}}
\right)\cdot\tilde{\nabla} \tilde{n} = D l \tilde{n}^{1/3} (\tilde{c}-
\tilde{c}_{e})\, H(\tilde{c}-\tilde{c}_{e}), \quad \mbox{where}\quad l=(48
\upi^2 v)^{1/3},   \label{4}
\end{eqnarray}
and $H(x)$ is the Heaviside unit step function. If $\tilde{c}<\tilde{c}_{e}$ (the vapour
concentration far from the droplets is smaller than the equilibrium concentration at the 
droplet surface), the vapour does not condense and the droplets do not grow. Eq.\ (\ref{4})
is the equivalent of Eq.\ (\ref{1}) for the condensate ($\rho\, (n-n_{*})$ instead of 
$\rho$ in Eq.\ (\ref{1})) accounting for the condensate source term. Eq.\ (\ref{4}) 
states that the steady growth of condensate due to gas advection and thermophoresis (left hand
side term) equals the growth of condensate due to vapour condensation on the particles (right 
hand side term) when both terms are divided by the number density of droplets $\rho$.

For the relatively large solid particle sizes we consider (about 1 micron), the equilibrium 
number density for which vapour coexists with a droplet is very close to the equilibrium 
number density for which vapour coexists with a half-space of liquid (assuming that the 
interphase is planar). The latter is given by the Clausius-Clapeyron relation, which for the 
case of an incompressible carrier gas, is
\begin{eqnarray}
{\tilde{c}_{e}\over\tilde{c}_{\infty}} = {\tilde{T}_{d}\over\tilde{T}}\, \exp
\left({\Lambda\over k_{B}\tilde{T}_{d}}- {\Lambda\over k_{B}\tilde{T}}\right).  
\label{5}
\end{eqnarray}
Here $\tilde{c}_{\infty}$ is a reference vapour density, $\Lambda$ is the heat of 
vaporization and $\tilde{T}_{d}$ is the {\em dew point temperature} at which 
$\tilde{c}_{\infty}=\tilde{c}_{e}$ in the absence of flow. In the presence of flow,
the dew point temperature changes and to determine its shift is part of the problem we have
to solve. 

If we have initially vapour at density $\tilde{c}_{\infty}$ and temperature $\tilde{T
}_{\infty}>\tilde{T}_{d}$, and lower the temperature below $\tilde{T}_{d}$, the 
vapour density $\tilde{c}_{\infty}$ is supersaturated. Then the solid particles carried by 
the gas act as condensation centers and the vapour can condense on them forming large 
droplets -- so large that capillary (Kelvin) effects are negligible. We consider only 
heterogeneous condensation of vapour on solid particles, thereby ignoring possible 
homogeneous condensation of vapour into droplets. The vapour follows the carrier gas 
flow and we neglect the Soret effect, \cite{CR88}\footnote{The solution of more detailed 
models (for example in OVD) show that changes due to the Soret effect are relatively small, 
\cite{fil03}; see also \cite{pedro} for a case in which the Soret effect plays an important 
role.}. Then the balance equation for the vapour number density is
\begin{eqnarray}
\left(\mathbf{\tilde{u}}\cdot\tilde{\nabla} - D\tilde{\Delta}\right)\tilde{c} 
= - D l \tilde{\rho} \tilde{n}^{1/3}(\tilde{c}-\tilde{c}_{e})\, H(\tilde{c}-
\tilde{c}_{e}).      \label{6}
\end{eqnarray}
Note that minus the right hand side of this equation equals that of Eq.\ (\ref{4}) times 
$\tilde{\rho}$; the negative sign occurs because a source for the condensate appears as a 
sink for the vapour. Then we can rewrite (\ref{6}) as
\begin{eqnarray}
\left(\mathbf{\tilde{u}}\cdot\tilde{\nabla} - D\tilde{\Delta}\right)\tilde{c} 
= \tilde{\rho} \left(\mathbf{\tilde{u}} - \alpha\nu{\tilde{\nabla}\tilde{T}
\over\tilde{T}}\right)\cdot\tilde{\nabla} \tilde{n}.      \label{6bis}
\end{eqnarray}

The temperature equation is 
\begin{eqnarray}
\mathbf{\tilde{u}}\cdot\tilde{\nabla}\tilde{T} = \kappa\tilde{\Delta}
\tilde{T},    \label{7}
\end{eqnarray}
where $\kappa$ is the thermal diffusivity. In this equation, we have ignored the Dufour 
effect, \cite{pedro}, and also the effect of the latent heat of condensation because the 
vapour mass fraction is very small compared to that of the carrier gas. Lastly, we need 
the equation for the velocity field, but we will not specify it for the time being because our 
theory can be applied to different flow fields. 

The boundary conditions for our problem are as follows. The temperature at infinity is 
$\tilde{T}_{\infty}$ and it is $\tilde{T}_{w}<\tilde{T}_{d}<\tilde{T}_{\infty}$ at 
the wall. Since $\tilde{T}_{w}<\tilde{T}_{d}$, we expect the vapour to have condensed on 
the cold wall and the vapour at the wall to be in local equilibrium with the liquid coating it. Thus 
$\tilde{c}=\tilde{c}_{e}$ at the wall. At infinity, the vapour density and droplet density are 
$\tilde{c}_{\infty}$ and $\tilde{\rho}_{\infty}$, respectively. At some 
distance from the wall, there is an interface between the condensation region where vapour 
condenses on the solid particles and coats them, and the outer region at a higher temperature 
where the particles are dry. To locate this {\em dew point interface} $\Gamma$ is part of the 
problem. At $\Gamma$, $\tilde{n}=n_{*}$, $\tilde{c}=\tilde{c}_{e}(\tilde{T}_{*})$ (from now
on, the asterisk will identify magnitudes at the dew interface), and the normal derivative of 
$\tilde{c}$ is continuous. 
Note that the dew point temperature at $\Gamma$ will be different from the dew point 
temperature in absence of flow, $\tilde{T}_{d}$. In short, the boundary conditions are:
\begin{eqnarray}
&& \tilde{T}=\tilde{T}_{\infty},\quad\tilde{c}=\tilde{c}_{\infty}, \quad\tilde{
\rho}=\tilde{\rho}_{\infty},\quad\mbox{at infinity,}\label{8}\\
&& \tilde{T}=\tilde{T}_{w},\quad\tilde{c}=\tilde{c}_{e}(\tilde{T}_{w}),
\quad\mbox{at the wall,}\label{9}\\
&& \tilde{n}=n_{*}, \quad\tilde{c}=\tilde{c}_{e}(\tilde{T}_{*}),\quad \left.
\left.\mathbf{n}\cdot\tilde{\nabla}\tilde{c}\right|_{\Gamma-} = \mathbf{n}
\cdot\tilde{\nabla}\tilde{c}\right|_{\Gamma+}, \quad\mbox{at }\Gamma.   
\label{10}
\end{eqnarray}
Assuming that we have calculated the carrier gas flow, $\mathbf{\tilde{u}}(\mathbf{\tilde{
x}})$, in principle we have enough boundary conditions to determine $\tilde{T}$, $\tilde{c}$, 
$\tilde{\rho}$, $\tilde{n}$ and $\Gamma$: 
\begin{itemize}
\item We solve the elliptic equation (\ref{7}) for $\tilde{T}$ with one condition at 
infinity and another at the wall, and the first order equation (\ref{2}) for $\rho$ with one 
boundary condition at infinity. 
\item For a given location of $\Gamma$, the first order equation (\ref{4}) for $\tilde{n}$ 
in the condensation region has one boundary condition at $\Gamma$. The elliptic equation 
(\ref{6}) for $\tilde{c}$ has Dirichlet boundary conditions (\ref{8}) at infinite and 
$\tilde{c}=\tilde{c}_{e}(\tilde{T}_{*})$ at $\Gamma$. Similarly, the equation for 
$\tilde{c}$ in the condensation region satisfies (\ref{9}) at the wall and 
$\tilde{c}=\tilde{c}_{e}(\tilde{T}_{*})$ at $\Gamma$. 
\item Given an arbitrary location of $\Gamma$, the two elliptic problems for $\tilde{c}$ 
are solved inside and outside the condensation region. Then the location of $\Gamma$ is
changed until the additional condition (\ref{10}) that the normal derivative of $\tilde{c}$ 
is continuous at $\Gamma$ is satisfied. This determines the position of the dew point interface. 
\end{itemize}

Note that the vapour concentration $\tilde{c}_{*}$ at the dew point interface is smaller 
than $\tilde{c}_{\infty}$ because the condensation region is a vapour sink and the 
diffusion causes a $\tilde{c}(\tilde{x})<\tilde{c}_{\infty}$ deficit even in the dry 
region outside the condensation region. Since $\tilde{c}_{*}=\tilde{c}_{e}(\tilde{T
}_{*})$ and $\tilde{c}_{\infty}=\tilde{c}_{e}(\tilde{T}_{d})$, we have $\tilde{c
}_{e}(\tilde{T}_{*})<\tilde{c}_{e}(\tilde{T}_{d})$. As $\tilde{c}_{e}(\tilde{
T})$ is an increasing function, we obtain $\tilde{T}_{*}<\tilde{T}_{d}$: due to the 
flow, the temperature at the dew point interface $\Gamma$ is lower than the dew point 
temperature in the absence of flow, $\tilde{T}_{d}$.

\section{Stagnation point flow}
\label{sec:stagnation}
\begin{figure}
\begin{center}
\includegraphics[width=10cm]{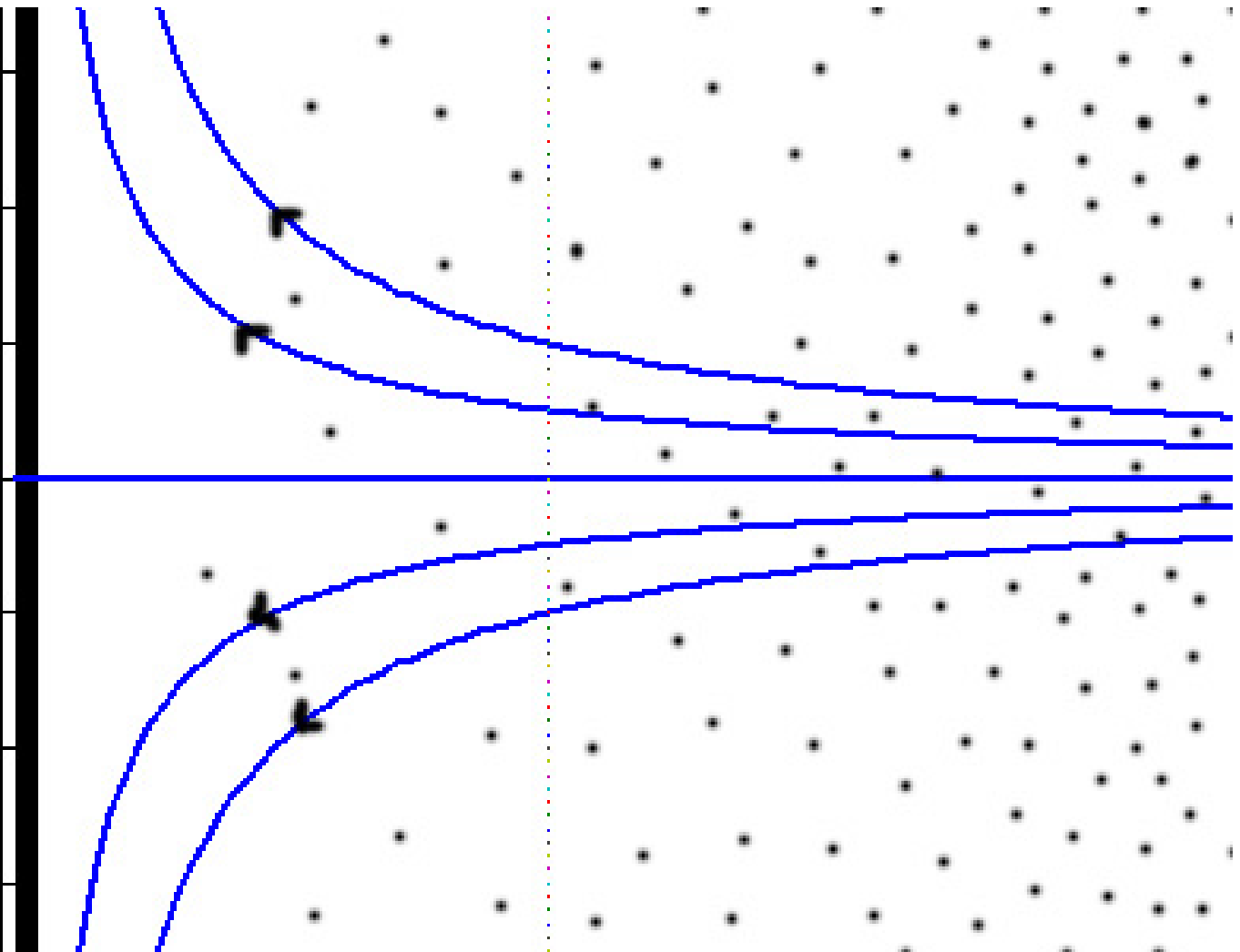}
\caption{Sketch of stagnation point flow with particles representative of vapour 
concentration.}
\label{fig0}
\end{center}
\end{figure}

As an example, consider the dew point shift in a Hiemenz stagnation point flow in the half 
space $\tilde{x}>0$ depicted in Figure \ref{fig0}, \cite{Hiemenz}. There is a solid wall 
at $\tilde{x}=0$ and the $\tilde{x}$ - velocity of the incoming flow is asymptotic to 
$-\gamma\tilde{x}$ as $\tilde{x}\to +\infty$, with a given strain rate $\gamma$. The 
boundary layer thickness is $l_{b}=\sqrt{\nu/\gamma}$, which we shall adopt as the unit of 
length. Then the unit of velocity is $\nu/l_{b}=\sqrt{\gamma\nu}$. We shall adopt 
$\tilde{c}_{\infty}$, $\tilde{\rho}_{\infty}$, $n_{*}$ and $\tilde{T}_{\infty}$ 
as the units of vapour density, droplet density, $n$ and temperature, respectively. Their values 
are given in table \ref{typicalvalues}.
\begin{table}%[ht]
\begin{center} \footnotesize
\begin{tabular}{@{}ccccccccl@{}}
% \hline
{$\tilde{T}_\infty$} & {$\tilde{T}_d$} & {$\tilde{c}_\infty$} & {$\tilde{
\rho}_\infty$} & {$n_*$} & {$l_{b}$} & {$\nu/l_{b}$} & {$a_{*}$} & {$v$}\\[3pt]
(K) & (K) & (cm$^{-3}$) & (cm$^{-3}$) & (--) & (mm) & (cm/s) & ($\mu$m) & cm$^3$\\
& & & & & & & & \\
%
%--&-- & -- & --  & $\displaystyle{4\pi a_{*}^3 \over 3v}$ & -- &-- &-- &--\\
%
%& & & & & & & & \\
%
1713 & 1400 & $1.9\times 10^{13}$ & $10^4$ & $4.72\times 10^{10}$ & 6.26 &
0.24 & 1 & $8.87\times 10^{-23}$ \\
% \hline
\end{tabular}
\end{center}
\caption{Typical parameters for heterogeneous condensation of Na$_{2}$SO$_{4}$ vapours
in air, \cite{CR88}.}
\label{typicalvalues}
\end{table}

The dimensionless $x$ component of the velocity is a function of $x$, denoted by $-u(x)$, 
$u>0$, whereas the dimensionless $y$ component of the velocity is $u'(x)\, y$. (Here and in 
the rest of the paper, $f'(x)$ means $df/dx$). Hence $u(x)$ is the parameter free solution of the 
well-known Hiemenz boundary value problem of stagnation in plane flow, \cite{Hiemenz}:
\begin{eqnarray}
&& u''' + u\, u'' + 1 - u'^2 = 0, \quad x>0,  \label{11}\\
&& u(0)= u'(0)=0,\quad u'(+\infty)=1. \label{12}
\end{eqnarray}
In nondimensional units, Eq.\ (\ref{7}) becomes
\begin{eqnarray}
&& T'' + \mbox{Pr}\, u\, T' = 0, \quad x>0,  \label{13}\\
&& T(0)= T_{w}={\tilde{T}_{w}\over\tilde{T}_{\infty}},\quad T(+\infty)=1, 
\label{14}
\end{eqnarray}
where Pr$=\nu/\kappa$ is the Prandtl number (which is 0.7 for air). Equations 
(\ref{2}), (\ref{4}) - (\ref{6}) with the boundary conditions (\ref{8}) - (\ref{10})
become
\begin{eqnarray}
&& \left(u+\alpha {T'\over T}\right)\rho' = -\alpha\,\rho\left({T'\over T}
\right)',\quad x>0, \label{15}\\
&&\rho(+\infty)= 1, \label{16}\\
&&\left(u+\alpha {T'\over T}\right)n' = - N\, n^{1/3}(c-c_{e}), \quad 0<x<x_{*}
\label{17}\\
&& n(x_{*})=1, \label{18}\\
&&c_{e}(x) = {T_{d}\over T(x)}\, \exp\left[{1\over \epsilon}\left({1\over T_d}
- {1\over T(x)}\right)\right],  \label{19}\\
&& c'' + \mbox{Sc}\, u\, c' = R\, \rho\, n^{1/3}(c-c_{e}), \quad 0<x<x_{*},   
\label{20}\\
&& c(0)=c_{e}(0), \quad c(x_{*})= c_{e}(x_{*}),\label{21}\\
&& c'' + \mbox{Sc}\, u\, c' = 0, \quad x>x_{*},   \label{22}\\
&& c(x_{*})= c_{e}(x_{*}), \quad c'(x_{*}-)= c'(x_{*}+), \quad c(+\infty)=1,\label{23}
\end{eqnarray}
where $x_{*}$ is the location of the dew point interface $\Gamma$, and
\begin{eqnarray}
\epsilon = {k_{B}\tilde{T}_{\infty}\over\Lambda},\quad \mbox{Sc}={\nu\over D},
\quad R={\nu l\tilde{\rho}_{\infty}n_{*}^{1/3}\over \gamma}, \quad
 N={D\tilde{c}_{\infty}l\over\gamma n_*^{2/3}} = {\tilde{c}_{\infty}\over\tilde{
 \rho}_{\infty}n_{*}\mbox{Sc}}\, R.  \label{24}
\end{eqnarray}
Here $l =(48\pi^2 v)^{1/3}$ and $v$ is the molecular volume. Note that $c_{e}$ 
given by the nondimensional version of the Clausius-Clapeyron relation (\ref{5}) is a 
function of $T$, and we are using the notation $c_{e}(x)=c_{e}(T(x))$ in Eq.\ (\ref{19}). 
Using (\ref{20}), we can rewrite (\ref{17}) as
\begin{eqnarray}
\left(u+\alpha {T'\over T}\right)n' = - \frac{N}{R\rho}\, (c'' + \mbox{Sc}\, u\, 
c'), \quad 0<x<x_{*},  \label{17bis}
\end{eqnarray}
which is analogous to (\ref{6bis}). Defining $U=u+\alpha\, T'/T$, we can use $n(x_{*})
=1$ and integrate (\ref{17bis}) to obtain
\begin{eqnarray}
n(x) = 1+ \frac{N}{R}\int_{x}^{x_{*}} \frac{c'' + \mbox{Sc}\, u\, c'}{\rho U}\, 
dx, \label{18bis}
\end{eqnarray}
which, integrated by parts yields
\begin{eqnarray}
n(x) = 1+ \frac{N}{R}\left[\frac{c'_{*}}{\rho_{*}U_{*}}- \frac{c'(x)}{\rho(x) 
U(x)} + \int_x^{x_{*}}\frac{\mbox{Sc}\, uU+u'}{\rho U^2}\, c'\, dx\right], 
\label{19bis}
\end{eqnarray}
due to (\ref{15}). Note that Equations (\ref{17bis}) - (\ref{19bis}) do not depend on
the model we use to describe vapour condensation on droplets.

In the limit as $R\to\infty$, $c\sim c_{e}$ for $0<x<x_{*}$ according to (\ref{20}), and 
then (\ref{17bis}) yields the approximate value of $n$ inside the condensation region. Note 
that the length $\delta$ over which $c$ decays to $c_{e}$ according to Eq.\ (\ref{20}) is 
found from a dominant balance between $(c-c_{e})''$ and the right hand side of (\ref{20}):
$$
(c-c_{e})''=\frac{[c-c_{e}]}{\delta^2}= R\, [\rho]\, [c-c_{e}]\Longrightarrow 
\delta=\frac{1}{\sqrt{R\, [\rho]}}.$$
where we have set the representative scale of $n$ as $[n]=1$. If we take $[\rho]=\rho(x_{*})
\equiv \rho_{*}$ as the scale of $\rho$, then $c$ decays to $c_{e}$ on a length given by 
\begin{eqnarray}
\delta={1\over\sqrt{R\rho_{*}}},   \label{43}
\end{eqnarray}
which goes to zero as $R\to +\infty$. In practice, $\rho_{*}$ is close to 1, and therefore 
$R^{-1/2}$ measures the dimensionless length over which $c$ decays to $c_{e}$. This length 
is just the width of the condensation layer in which there is supersaturation and therefore the
vapour condenses on droplets. If $\delta\ll x_{*}$, the condensate arriving to the wall is 
mostly due to the arrival of solid particles coated with liquid, whereas for larger $\delta$, 
direct condensation of vapour on the wall is important. Thus the parameter $R$ gives an idea 
of the vapour scavenged by condensation on solid particles: the larger $R$ is, the more 
vapour condenses on particles.

\begin{table}%[ht]
\begin{center} \footnotesize
\begin{tabular}{@{}cccccccccccl@{}}
% \hline
-- & & {$\alpha$} & {$\epsilon$} & {Pr} & {Sc} & {$R$}& {$N$} & {$\delta_{e}$} 
& {$\delta$} & {$\mu$} & $T_{w}$\\[3pt]
% \hline
%
-- &&  --
    & $\displaystyle{k_{B}\tilde{T}_\infty \over \Lambda}$ 
    & $\displaystyle{\kappa \over D}$
    & $\displaystyle{\nu \over D}$ 
    & $4\upi a_{*}\tilde{\rho}_{\infty} l^2_{b}$
    & $\displaystyle{4\upi a_{*} l^2_{b}\tilde{c}_\infty \over n_*\mbox{Sc}}$
    & $\displaystyle{\epsilon T^2_*\over T'_*}$
    &$\displaystyle{1\over\sqrt{R\rho_{*}}}$
    &$\displaystyle{\delta\over\delta_{e}}$&$\displaystyle{\tilde{T}_{w}\over\tilde{T}_{\infty}}$\\
% \hline
& & & & & & & & & & & \\
A & & 0.1 & 0.0515& 0.7 & 1.8 & 4.93 & 0.11 & 0.1572& 0.4559 &2.9001& 0.5838\\
 %\hline
B && 0.1 & 0.0515& 0.7 & 1.8 & 73.95 & 0.11 & 0.1572& 0.1177 &0.7487&  0.5838\\
 %\hline
C && 0.1 & 0.0515& 0.7 & 1.8 & 739.5 & 1.65 & 0.1572& 0.0372 &0.2366& 0.5838 \\
 %\hline
\end{tabular}
\end{center}
\label{parameters}
\caption{Dimensionless parameters for A: $l_{b}=6.26$ mm, $\tilde{\rho}_{\infty}=
10^4$ cm$^{-3}$, B: $l_{b}=6.26$ mm, $\tilde{\rho}_{\infty}=1.5\times 10^5$ 
cm$^{-3}$, and C: $l_{b}=2.42$ cm, $\tilde{\rho}_{\infty}=10^5$ cm$^{-3}$. The
values of $\delta_{e}$, $\delta$ and $\mu$ (which depend on $T_{w}$) have been 
calculated using the 0-CL theory of Section \ref{sec:equilibrium}. Other values as in table 
\ref{typicalvalues}.}
\end{table}
Representative values for the parameters (\ref{24}) are given in table 2 for
Na$_{2}$SO$_{4}$ vapours in air as in \cite{CR88}. The parameter $R$ can be rewritten as
 $$R=\tilde{\rho}_{\infty}n_{*}^{1/3}ll_{b}^2=\tilde{\rho}_{\infty}(48\upi^2 v\, 
 n_{*})^{1/3} l_{b}^2= 4\upi a_{*}\tilde{\rho}_{\infty}l^2_{b},
 $$ 
 where $l_{b}=\sqrt{\nu/\gamma}$ is the width of the Hiemenz boundary layer. For 
 Na$_{2}$SO$_{4}$, whose mass density is 2.66 g/cm$^3$, we obtain a molecular volume $v= 
 8.87\times 10^{-23}$ cm$^3$ using a molecular weight of 142. A solid particle of radius 
 $a_{*}=1\,\mu$m has a volume equivalent to a liquid drop with $n_{*}= 4.72\times 10^{10}$ 
 molecules. The prefactor $R/l_{b}^2=4\pi a_{*}\tilde{\rho}_{\infty}$ is 12.57 
 cm$^{-2}$ and $R=1$ for a boundary layer which is 2.82 mm thick. For a typical boundary 
 layer experiment with a displacement thickness of 11 mm, $l_{b}=1.1/1.72= 0.63$ cm, $R= 
 4.93$ and $N=0.11$, which corresponds to entry A in table 2. Entry B corresponds to a $R$ 
that is 15 times larger than that in entry A. Entry C in table 2 corresponds to $R$ and $N$ 
that are 150 and 15 times larger than those in entry A, respectively. 

We have used three different boundary layer widths and solid particle densities to illustrate 
different asymptotic theories for the condensation layer based on the fact that $\epsilon$ is 
typically small. In this case, the Clausius-Clapeyron relation (\ref{19}) may be written as
$$\frac{c_{e}(x)}{c_{*}} = \frac{T_{*}}{T(x)}\,\exp\left(\frac{T(x)-T_{*}}{\epsilon
T(x)T_{*}}\right)\approx\exp\left(-\frac{x_{*}-x}{\delta_{e}}\right),$$
provided
\begin{equation}
\delta_{e}= \frac{\epsilon T^2_{*}}{T'_{*}}.\label{26bis}
\end{equation}
The equilibrium vapour concentration decays exponentially fast in a dimensionless length
$\delta_{e}$ given by (\ref{26bis}). If $\delta\ll\delta_{e}$, the vapour density
decays to $c_{e}(x)$ much faster than the equilibrium vapour density decays to zero, even
if $\delta_{e}$ or $\delta$ are not particularly small. The simplest asymptotic theory we
can propose is based on assuming $\delta=0$ and $c(x)=c_{e}(x)$ inside the dew 
surface, $0<x<x_{*}$. This is the simplified equilibrium theory (or zero-width CL theory,
0-CL) already studied by \cite{CR89}. If $\delta\ll\delta_{e}$, there is a thin CL of 
width $\delta$ inside the dew surface. A first correction to considering a zero-width CL as it 
does the 0-CL theory may follow from assuming that the CL is so narrow that $\rho$, $T$ 
and $u$ do not differ appreciably from their values at $x_{*}$. This $\delta$-wide CL 
theory ($\delta$-CL theory) corrects the simplified equilibrium theory even for relatively 
small values of the scavenging parameter $R$ as in the case A of Table 2; cf.\ Section 
\ref{sec:condlayer}. For large values of $R$ as in case C of Table 2, $\delta\to 0$, and the 
length scale on which $c-c_{e}$ varies is much smaller than $x$. Then a method of multiple 
scales may provide an accurate description of the CL as shown in Section \ref{sec:attached}.
Our considerations on the validity of the 0-CL theory can be repeated for a general boundary
layer flow.

\subsection{Deposition at the wall}
\label{sec:flux}
At the wall, both vapour is directly condensed and droplets are deposited. The respective 
fluxes of condensate at the wall are 
\begin{eqnarray}
&& - \tilde{J}_{v}= D\,\tilde{c}'(0)= {D\tilde{c}_{\infty}\over l_{b}}\, c'(0), 
\label{74}\\
&& - \tilde{J}_{c}= \tilde{\rho}(0)\,\tilde{U}(0)\, [n(0)-n_{*}] = {\nu\tilde{\rho}_{
\infty} n_{*}\over l_{b}}\, {\alpha T'(0)\rho(0)\, [n(0)-1]\over T(0)}.  \label{75}
\end{eqnarray}
Choosing $\nu\tilde{c}_{\infty}/l_{b}$ as the unit of flux, the nondimensional fluxes 
are
\begin{eqnarray}
J_{v}&=& {c'(0)\over\mbox{Sc}},\label{76bis}\\
J_{c}&=& {R\over N\,\mbox{Sc}}\, {\alpha\rho(0)\,
T'(0)\, [n(0)-1]\over T(0)}= {\rho(0)U(0)\over\mbox{Sc}}\,\int_{0}^{x_{*}}{c'' 
+ \mbox{Sc}\, u\, c'\over\rho U}\, dx,  \label{76}
\end{eqnarray}
where we have omitted the minus signs and used (\ref{18bis}). The total flux of condensate 
at the wall is
\begin{eqnarray}
J\equiv J_{v}+J_{c}= {c'(0)\over\mbox{Sc}} + {R\over N\,\mbox{Sc}}\,\rho(0)\, 
U(0)\, [n(0)-1].  \label{77}
\end{eqnarray}
Using Eq.\ (\ref{19bis}), this equation becomes
\begin{eqnarray}
J= {\rho(0)U(0)\over\mbox{Sc}}\left[\frac{c'_{*}}{\rho_{*}U_{*}} + \int_0^{
x_{*}}\frac{\mbox{Sc}\, uU+u'}{\rho U^2}\, c'\, dx\right].        \label{77bis}
\end{eqnarray}

\subsection{Formulas for the temperature profile and for the vapour concentration profile
in the dry region}
The temperature profile is a solution of (\ref{13}) - (\ref{14}) and the vapor concentration
satisfies in the dry region, $0<x<x_{*}$, a similar equation (\ref{22}) with boundary 
conditions (\ref{23}). The following argument\footnote{Yossi Farjoun, private 
communication.} provides an efficient way to solve these shooting problems. Note that $c-1$ 
is a solution of Eq.\ (\ref{22}) with zero value at $x=+\infty$ and
that any multiple of this solution also becomes zero at infinity (although of course it takes
on a different value at the other boundary). Thus we consider the following universal problem
\begin{eqnarray}
\psi'' +\mbox{Sc}\, u \psi'= 0,\quad \psi(0)=1,\quad \psi(+\infty)=0. \label{31}
\end{eqnarray}
The unique solution of this problem yields the solution of Eq.\ (\ref{22}) with boundary 
conditions $c(x_{*})=c_{*}$ and $c(+\infty)=1$:
\begin{eqnarray}
c(x)= 1 + {c_{*}-1\over\psi_{*}}\,\psi(x). \label{32}
\end{eqnarray}
Similarly, 
\begin{eqnarray}
&& T(x)= 1 + (T_{w}-1)\,\Phi(x), \label{t1}\\
&& \Phi'' +\mbox{Pr}\, u \Phi'= 0,\quad \Phi(0)=1,\quad \Phi(+\infty)=0,\label{t2}
\end{eqnarray}
solves (\ref{13}) with boundary conditions (\ref{14}), $T(0)=T_{w}$, $T(+\infty)=1$. 
Analytic expressions for $\psi$ and $\Phi$ are found by direct integration of the linear 
equations (\ref{22}) and (\ref{13}) with boundary conditions $\Phi(0)=\psi(0)=1$, 
$\Phi(+\infty)=\psi(+\infty)=0$: 
\begin{eqnarray}
\psi(x)= {\int_{x}^\infty e^{-\mbox{\footnotesize Sc}\, \int_{0}^y u dx'}dy
\over\int_{0}^\infty e^{-\mbox{\footnotesize Sc}\, \int_{0}^y u dx'}dy}, 
\label{34}\quad
\Phi(x)={\int_{x}^\infty e^{-\mbox{\footnotesize Pr}\, \int_{0}^y u dx'}dy
\over\int_{0}^\infty e^{-\mbox{\footnotesize Pr}\, \int_{0}^y u dx'}dy}.
\end{eqnarray}
If the dew point interface coincides with the wall, $x_{*}=0$, $\psi_{*}=\psi(0)=1$ and 
$c(0)=c_{e}(0)$. Then Eqs.\ (\ref{32}) and (\ref{34}) yield 
\begin{eqnarray}
c(x)= c_{e}(0)+{[1-c_{e}(0)]\,\int_{0}^x\exp(-\mbox{Sc}\int_{0}^y u dx')\, dy
\over\int_{0}^\infty \exp(-\mbox{Sc}\int_{0}^y u dx')\, dx}, \,\label{78bis}
\end{eqnarray}
cf.\ \cite{CR88}. The deposition flux due to droplets is zero and $J=J_{v}$ is
\begin{eqnarray}
 J = {1-c_{e}(0)\over\mbox{Sc}\,\int_{0}^\infty \exp(-\mbox{Sc}\int_{0}^y u 
dx')\, dx}.  \label{78}
\end{eqnarray}

\subsection{Numerical method}
We have used the following numerical procedure to find $x_{*}$. We numerically solve the 
universal shooting problems (\ref{31}) and (\ref{t2}) with sufficient accuracy. Similarly, 
we numerically solve (\ref{11}) - (\ref{12}) and (\ref{15}) - (\ref{16}) to determine 
$u$ and $\rho$ with sufficient accuracy. Then the temperature profile (\ref{t1}), $u$ and 
$\rho$ are known.
\begin{itemize}
\item We start from a trial value of $x_{*}$. 
\item We numerically solve Eq.\ (\ref{20}) for $x<x_{*}$ with initial conditions $c(x_{*})
=c_{*}=c_{e}(x_{*})$ and $c'(x_{*})=c'_{*}= (c_{*}-1)\psi'_{*}/\psi_{*}$ obtained 
from Eq.\ (\ref{32}).
\item We compare the value $c(0)$ given by the numerical solution with $c_{e}(0)$. If they
are not equal, we change the value of $x_{*}$ and repeat the procedure until we find
$c(0)=c_{e}(0)$.
\end{itemize}

\section{Zero width condensation layer: Simplified equilibrium model}
\label{sec:equilibrium}
In this Section, we shall assume that the relaxation to local equilibrium between vapour and
droplets in the condensation region is so fast that the vapour density in that region equals
$c_e$ given by the Clausius-Clapeyron relation (\ref{5}) with a temperature field obeying
Eq.\ (\ref{7}). Then the width of the CL is zero (0-CL theory). This simplified 
equilibrium model was introduced and studied by \cite{CR89}. Here we revisit the model 
and give approximate formulas for the dew point shift from $T_{d}$ due to the flow, for the 
location of the dew point interface, for the deposition rates and for the maximum wall 
temperature having nonzero $J_{c}$. In later sections, we shall determine how well the 0-CL 
theory approximates the solution of the complete thermophysical model in Section 
\ref{sec:model}. 

The validity of the simplified equilibrium model requires $\delta\to 0$, where $\delta$
is the dimensionless length over which $c-c_{e}\to 0$ as indicated in Section \ref{sec:stagnation}
for the case of stagnation point flow. 
Similar considerations apply in the case of a general boundary layer flow. Adopting the 
same units as in Section \ref{sec:stagnation} to nondimensionalize the governing equations 
of the model, the considerations made there apply to the general case for which $l_{b}$ is a 
characteristic length associated with the carrier gas flow. Near the dew interface $\Gamma$, 
the Clausius-Clapeyron relation (\ref{19}) indicates that $c_{e}(\mathbf{x})$ decays rapidly
as we move inside the condensation region with the dimensionless length constant $\delta_{e}$
given by (\ref{26bis}), in which 
%$${c_{e}\over c_{*}} = {T_{*}\over T}\,\exp\left({T-T_{*}\over\epsilon TT_{*}}
%\right)\Longrightarrow {c'_{e}\over c_{e}} = {(1-\epsilon T_*)\, T'_*\over\epsilon T_*^2},$$
%which yields
%\begin{eqnarray}
%&&\mathbf{n}\cdot\nabla c_* = (1-\epsilon T_*)\, {c_*\over\delta_e}, \label{25}\\
%&&\delta_e = {\epsilon\, T_*^2\over\mathbf{n}\cdot\nabla T_*}.  
%\label{26}
%\end{eqnarray}
 the asterisk denotes evaluation at $\Gamma$, $T'=\mathbf{n}\cdot\nabla T$ 
and $\mathbf{n}$ is the outer normal to $\Gamma$ pointing away from the wall. 
$\delta_{e}$ expressed in terms of dimensional parameters is
\begin{eqnarray}
\delta_e = {k_B\tilde{T}_*^2\over\Lambda l_{b}\mathbf{n}
\cdot\tilde{\nabla}\tilde{T}_*},  \label{27}
\end{eqnarray}
typically, $\delta_e\ll 1$. For the numerical values of the parameters employed in Section
\ref{sec:model}, we find $\delta_{e}= 0.1572$.

Using $l_{b}$ and $\nu/l_{b}$ as the units of length 
and velocity, the nondimensional equation for the vapour density in the region between 
$\Gamma$ and infinity (the {\em dry} region) is
\begin{eqnarray}
\Delta c - \mbox{Sc}\, \mathbf{u}\cdot\nabla c = 0.   \label{28}
\end{eqnarray}
 The boundary condition far from the dew point interface (at infinity) is $c\to 1$ (in 
 nondimensional units), whereas $c_{*}=c_{e}(T_{*})$ on the dew point surface $\Gamma$. 
 In typical geometries with a known $\Gamma$, this boundary value problem (\ref{28}) is 
 well-posed and $c$ and its normal derivative are functionals of $\Gamma$. To determine the 
 dew point interface, we have to use that the normal derivative of $c$ is continuous on 
 $\Gamma$. Using the Clausius-Clapeyron formula (\ref{19}), this condition becomes
\begin{eqnarray}
(1-\epsilon T_*)\, {T_{d}\over T_*}\,\exp\left[{1\over \epsilon}\left({1\over 
T_d} - {1\over T_*}\right)\right] =  {\epsilon\, T_*^2\,\mathbf{n}\cdot
\nabla c_*\over\mathbf{n}\cdot\nabla T_*}.  \label{29}
\end{eqnarray}
The dew point interface is determined by solving Eq.\ (\ref{28}) with boundary conditions 
$c=c_{e}$ on $\Gamma$ and $c=1$ at infinity for different $\Gamma$ until Eq.\ (\ref{29}) 
is satisfied.

In summary, using the simplified equilibrium model, the dew point interface is determined as
follows:
\begin{itemize}
\item The flow field $\mathbf{u}$ and the temperature field $T$ have been determined
beforehand and are considered known.
\item Eq.\ (\ref{28}) is solved with boundary conditions $c=c_{e}$ on $\Gamma$ and
$c=1$ at infinity for an arbitrary position of the interface $\Gamma$.
\item The position of $\Gamma$ is changed until (\ref{29}) holds. 
\end{itemize}
In what follows, we consider the dew point shift in a Hiemenz stagnation point flow of Section 
\ref{sec:stagnation}. 

\subsection{Dew point location}
Once Eqs.\ (\ref{11}) - (\ref{14}) are solved, Eq.\ (\ref{19}) yields the 
local equilibrium vapour density $c_{e}(x)$, which equals the vapour number density for $x<
x_{*}$. At the unknown position $x_{*}$, the vapour number density and its derivative are 
obtained from Eq.\ (\ref{19}):
\begin{eqnarray}
c_{*}= {T_{d}\over T_{*}}\,\exp\left[{1\over\epsilon}\left({1\over T_{d}}-
{1\over T_{*}}\right)\right],\quad  c'_{*}= (1-\epsilon T_{*}) {T'_{*} c_{*}\over
\epsilon T_{*}^2}, \label{30}
\end{eqnarray}
where $T_{*}=T(x_{*})$. We now solve Eq.\ (\ref{22}) for $x>x_{*}$ with initial 
conditions (\ref{30}) and calculate $c(+\infty)$. We keep changing $x_{*}$
until we obtain the correct boundary condition $c(+\infty)=1$ in Eq.\ (\ref{23}). 
In terms of the solution $\psi(x)$ of (\ref{32}), we can calculate directly 
\begin{eqnarray}
c'(x)= {c_{*}-1\over\psi_{*}}\,\psi'(x)\Longrightarrow c'_{*}= {c_{*}-1\over
\psi_{*}}\,\psi'_{*}. \label{33}
\end{eqnarray}
The location $x_{*}$ is found by equating $c'_{*}$ to the expression (\ref{30}). Inserting
the analytical solution (\ref{34}) in Eq.\ (\ref{33}), we get 
\begin{eqnarray}
{c'_{*}\over 1- c_{*}}= -\frac{\psi'_{*}}{\psi_{*}}={1\over\int_{x_{*}}^\infty 
e^{-\mbox{\footnotesize Sc}\,\int_{x_{*}}^y u dx'} dy}.    \label{35}
\end{eqnarray}
See Eq.\ (68) in \cite{CR88}. In this reference it is also proved that $x_{*}$ as
given by the simple equilibrium theory is always closer to the wall than the value given by
solving the full problem (\ref{15}) - (\ref{23}). 

The condition of continuity of $c'(x)$ at $x=x_{*}$ is satisfied if we insert (\ref{30}) in
(\ref{35}). Then the left hand side of (\ref{35}) can be rewritten as
$$\frac{(1-\epsilon T_{*}) T'_{*}}{\epsilon T_{*}^2\left(\frac{1}{c_{*}}-1
\right)}=\frac{(1-\epsilon T_{*}) T'_{*}}{\epsilon T_{*}^2\left(\frac{T_{*}}
{T_{d}}\,\exp\left[\frac{1}{\epsilon}\left(\frac{1}{T_{*}}-\frac{1}{T_{d}}
\right)\right]-1\right)}.
$$
Now inserting $T=1+(T_{w}-1)\,\Phi(x)$ in (\ref{35}), we obtain the following equation 
for $x_{*}$:
\begin{eqnarray}
\frac{\psi'_{*}}{\psi_{*}\Phi'_{*}}=\frac{[1-\epsilon + \epsilon (1-T_{w})
\Phi_{*}]\, (1-T_{w})}{\epsilon [1-(1-T_{w})\Phi_{*}]^2\left(\frac{1-(1-T_{w})
\Phi_{*}}{T_{d}}\,\exp\left[\frac{1}{\epsilon}\left(\frac{1}{1-(1-T_{w})
\Phi_{*}}-\frac{1}{T_{d}}\right)\right]-1\right)}.    \label{30u}
\end{eqnarray}

\subsection{Dew point shift}
\begin{figure}
\begin{center}
\includegraphics[width=10cm]{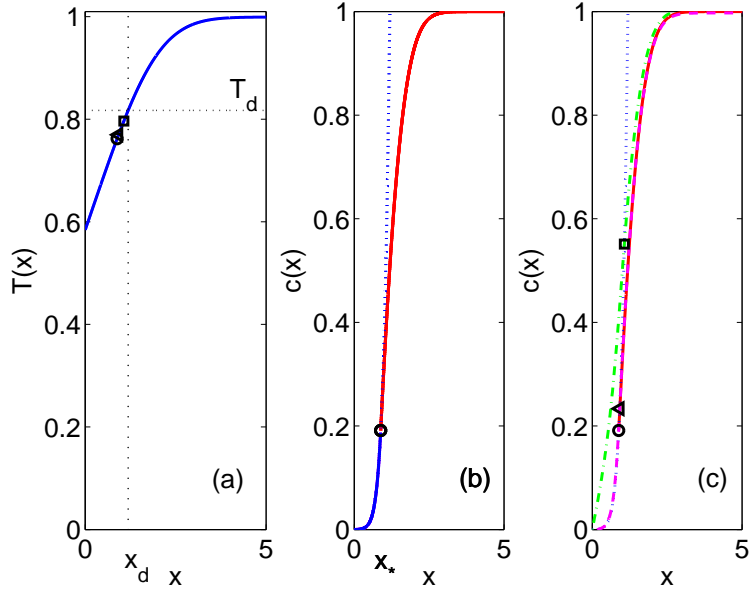}
\caption{(a) Nondimensional temperature ($T=\tilde{T}/\tilde{T}_{\infty}$) versus
distance to the wall ($x=\tilde{x}/l_{b}$). We have marked the dew point temperature 
$T_{d}$ in the absence of flow, the approximate value of the dew point interface $(x_{*},
T_{*})$ according to the simplified equilibrium model having a condensation layer of zero 
width (circle) and the result of a direct numerical calculation of that interface for entries A 
(square) and C (triangle) in table 2. For entry C in table 2, 
the result of direct numerical calculation is very close to the prediction of the asymptotic 
theory. (b) vapour density $c$ versus $x$ given by the approximate equilibrium model with the
location of the dew point interface marked by a circle. (c) Same as (b) but adding $c(x)$ as 
given by direct numerical calculation for entry C in table 2 (solid line; the 
triangle marks the dew point interface) and by direct numerical calculation for entry A in 
table 2 (dot-dashed line; the square marks the dew point interface). Here 
$T_{w}=0.5838$, $T_{d}=0.8173$.}
\label{fig1}
\end{center}
\end{figure}

For the parameter values of Section \ref{sec:stagnation}, the temperature and vapour density 
profiles are depicted in figure \ref{fig1}. The  dew point position turns out to be $x_{*}=
0.8815$, and the corresponding dimensionless temperature is $T_{*}= 0.7620$, i.e., 1305 K 
because $\tilde{T}_{\infty}= 1713$ K. We obtain a dew point shift $\tilde{T}_{*}-
\tilde{T}_{d}= -95$ K. An approximate formula can be obtained by rewriting Eq.\ 
(\ref{29}) as
\begin{eqnarray}
T_{*}- T_{d}= T_{d} T_{*}\epsilon\left[\ln\epsilon + \ln\left({T_*^3 c'_{*}
\over T_d T'_{*}}\right) - \ln(1-\epsilon T_*)\right].  \label{36}
\end{eqnarray}
As $\epsilon\to 0$, $T_{*}\to T_{d}$, so we can approximate $T_{*}\approx T_{d}$,
$T'_{*}\approx T'_{d}$ and $c'_{*}\approx c'_{d}$ in this formula to get
\begin{eqnarray}
T_{*}- T_{d}\approx T_{d}^2\epsilon\left[\ln\epsilon + \ln\left({T_d^2c'_{d}
\over T'_d}\right) - \ln(1-\epsilon T_d)\right].  \label{37}
\end{eqnarray}
$x_{d}$ is calculated by solving $T(x_{d})=T_{d}$. Then $c'_{d}=c'(x_{d})$. Eq.\ 
(\ref{37}) yields a dew point shift - 114 K, with a 13.6\% relative error. We see from
figure \ref{fig1} that the decrease of vapour concentration from $x=+\infty$ to $x=x_{*}$
is dramatic according to the simplified equilibrium model, from $\tilde{c}_{\infty}= 1.90
\times 10^{13}$ cm$^{-3}$ to $\tilde{c}_{*}= 0.1910\, \tilde{c}_{\infty} =0.363
\times 10^{13}$ cm$^{-3}$. The simplified equilibrium model is a good approximation for 
large values of $R$ and $N$ as in entry C of table 2. For moderate values of 
$R$ and $N$ as in entries A and B of table 2, the approximate theory overestimates 
the decrease of vapour concentration at the dew point interface.

\subsection{Maximum wall temperature at which there is a CL}
As $T_{w}$ increases, $x_{*}$ decreases until $x_{*}=0$. This marks the absence of a CL
of finite width. At the corresponding wall temperature, $T_{w,M}$, which is independent on 
the model we use to describe vapour condensation on droplets, $J_{c}=0$. At $T_{w,M}$,
$\Phi_{*}=\psi_{*}=1$ and the condition (\ref{30u}) provides the following equation
for $T_{w,M}$ as given by the 0-CL theory:
\begin{eqnarray}
\frac{\Phi'(0)}{\psi'(0)}=\frac{\epsilon T_{w,M}^2}{(1-T_{w,M})(1-\epsilon
T_{w,M})} \left\{
\frac{T_{w,M}}{T_{d}}\,\exp\left[\frac{1}{\epsilon}\left(\frac{1}{T_{w,M}}
-\frac{1}{T_{d}}\right)\right]-1\right\}.    \label{tmax}
\end{eqnarray}
For $T_{d}= 0.817$ (1400 K) and $\epsilon=0.0515$ (as in table 2), we obtain $T_{w,M}
=0.755$ (1293 K).

\subsection{Deposition at the wall}
Eq.\ (\ref{17bis}) holds in the condensation region, $0<x<x_{*}$ (and $x_{*}>0$ for
$T_{w}<T_{w,M}$), no matter which formula we use for the rate of vapour condensation 
on droplets, \cite{CR89}. The vapour deposition rate and the total deposition rate at the 
wall are given by inserting $c(x)=c_{e}(x)$ in Eqs.\ (\ref{76bis}) and (\ref{77bis}), 
respectively. Then $J_{c}=J-J_{v}$.

At $T_{w,M}$ given by (\ref{tmax}), $J_{c}=0$ and $J=J_{v}=c'_{e}(0)/$Sc. For $T_{w,
M}\leq T_{w}<T_{d}$, $J_{c}=0$ and the deposition rate is given by $J=[1-c_{e}(T_{w})]\, 
|\psi'(0)|/$Sc (here we are using $c_{e}$ as a function of $T$). For wall temperatures slightly 
below $T_{w,M}$, (\ref{76bis}), (\ref{76}) and (\ref{30u}) give $J= c'_{e}(0)+
c''_{e}(0)x_{*} + O(x^2_{*})$ with $x_{*}\sim (T_{w,M}-T_{w})/[(1-T_{w,M})\,
|\Phi'(0)|]$, as it follows from (\ref{19}). Here $c_{e}(x)= c_{e}(T(x))$ and its derivatives
are calculated for $T(0)=T_{w,M}$. This yields
\begin{eqnarray}
J\sim\frac{1-T_{w,M}}{\mbox{Sc}\epsilon T^2_{w,M}}|\Phi'(0)| c_{e}(T_{w,M})
\left[1-\epsilon T_{w,M} +\frac{1-4\epsilon T_{w,M}}{\epsilon T_{w,M}^2} \,
(T_{w,M}-T_{w})\right].   \label{nearby}
\end{eqnarray}

\section{Small-width condensation layer ($\delta$-CL)}
\label{sec:condlayer}
In this Section, we shall correct the 0-CL theory (simplified equilibrium theory) in the case 
$\delta\ll\delta_{e}\ll 1$ by assuming that the CL inside the dew surface is very thin
and detached from the wall. Then $\rho$, $T$ and $u$ can be approximated by their values
at $x_{*}$. The resulting $\delta$-CL theory should hold even for relatively small $R$ as 
in case A of table 2. 

%\subsection{Dew point location}
For stagnation point flow, it is convenient to work with the nondimensional equations 
(\ref{15}) - (\ref{23}). Using $C=c-c_{e}$, $R\rho_{*}=1/\delta^2$, and ignoring the 
convective term (because $c'\ll c''$ and $c'_{e}\ll c''_{e}$ if $\delta\ll\delta_{e}\ll 
1$), Eq.\ (\ref{20}) becomes
\begin{eqnarray}
C'' - {1\over\delta^2} n^{1/3}C = - c''_{e} = -{c_{*}\over\delta_{e}^2}\, 
e^{(x-x_{*})/\delta_{e}}. \label{44}
\end{eqnarray}
Eq.\ (\ref{44}) should be solved with boundary conditions
\begin{eqnarray}
C(x_{*}) = 0, \quad C(0)=0, \quad  c'_{e}(x_{*}) + C'(x_{*}) = c'_{*}. \label{45}
\end{eqnarray}
Here $c'_{*}$ is determined by solving Eqs.\ (\ref{22}) - (\ref{23}) in addition to 
(\ref{44}) - (\ref{45}). Using Eq.\ (\ref{30}) for the equilibrium vapour density, the last 
condition (\ref{45}) can be rewritten as 
\begin{eqnarray}
(1-\epsilon T_{*})\, {c_{*}\over\delta_{e}} + C'(x_{*}) = c'_{*}. \label{46}
\end{eqnarray}
Similarly, Eqs.\ (\ref{17}) - (\ref{18}) become
\begin{eqnarray}
n' = - {N\over U_{*}}\, n^{1/3}C , \quad n(x_{*})=1. \label{47}
\end{eqnarray}
A rough approximation to $n(x)$ is found as follows. We approximate $\rho U\approx 
\rho_{*}U_{*}$ and $c''+$Sc$\, u\, c'\approx c''$  in (\ref{18bis}), thereby obtaining
\begin{eqnarray}
n(x) = 1 + {N\over R\rho_{*}U_{*}}\, [c'_{*}-c'(x)], \label{48}
\end{eqnarray}
or, equivalently,
\begin{eqnarray}
&& {n(x)-1\over n_{l}} = 1 - {c'(x)\over c'_{*}} , \label{49}\\
&& n_{l} = {N\, c'_{*}\over R\rho_{*}U_{*}}= {D\, |\mathbf{n}\cdot
\tilde{\nabla}\tilde{c}_*|\over\tilde{\rho}_{*}\tilde{U}_{*}n_{*}}=
\frac{n_{\infty}}{n_{*}}. \label{50}
\end{eqnarray}
Here $n_{l}$ is the eventual number of vapour molecules per droplet in units of $n_{*}$. For 
the parameter values in table 2 corresponding to a dew point temperature $\tilde{T}_{d}=
1400$ K, $n_{l}= 0.044$ and $n_{\infty}=n_{l}n_{*}= 3.1\times 10^9$. The number of 
vapour molecules needed to cover a solid particle of radius $[3n_{*}v/(4\upi)]^{1/3}$ with 
a single layer of liquid is $n_{s}= 4 n_{*}^{2/3} = 5.2\times 10^7$. Thus $n_{\infty}/
n_{s}= 60$ layers of condensed vapour cover each solid particle in our numerical 
example and therefore our theory based on continuum diffusive growth of droplets yields 
consistent results.

\subsection{Vapour density profile and dew point location}
Eqs.\ (\ref{44}) and (\ref{47}) can be written as a boundary layer problem in a new 
variable $\xi=(x_{*}-x)/\delta_{e}$, with boundary conditions $C=0$ at $\xi=0$ ($x=
x_{*}$) and at $\xi=x_{*}/\delta_{e}\gg 1$ ($x=0$):
\begin{eqnarray}
&& \mu^2\, {d^2\over d\xi^2}( C+ c_{*}e^{-\xi}) - n^{1/3} C= 0,  \label{51}\\
&& \mu\, {dn\over d\xi} = {n_{l}\over c'_{*}\delta}\, n^{1/3} C,\label{52}
\end{eqnarray}
where 
\begin{eqnarray}
\mu = {\delta\over\delta_{e}},     \label{53}
\end{eqnarray}
is the ratio of lengths for $C$ and $c_{e}$ to decay to zero. Our numerical example shows 
that $n_{l}$ is small, Eq.\ (\ref{49}) implies that $n-1\to n_{l}$ as $\xi\to\infty$ and
Eq.\ (\ref{52}) with boundary condition $n(0)=1$ indicates that $0\leq n-1\leq n_{l}$, so 
that $n$ is always close to 1. If $n\approx 1$, Eq.\ (\ref{51}) becomes a linear problem:
\begin{eqnarray}
\mu^2\, {d^2\over d\xi^2}( C+ c_{*}e^{-\xi}) - C= 0,  \quad
C(0)=0, \quad C\left(\frac{x_{*}}{\delta_{e}}\right)=0, \label{54}
\end{eqnarray}
whose solution is 
\begin{eqnarray}
C &=&\frac{c_*\mu^2}{\mu^2-1} \left[\frac{e^{-x_{*}/\delta_{e}}\sinh\left(
\frac{\xi}{\mu}\right) -
\sinh\left(\frac{\xi-x_{*}/\delta_{e}}{\mu}\right)}
{\sinh\left(\frac{x_{*}}{\mu\delta_{e}}\right)}-e^{-\xi}\right] \nonumber\\
&=& \frac{c_*\delta^2}{\delta^2-\delta^2_{e}}\left[\frac{
\sinh\left(\frac{x}{\delta}\right)+ e^{-x_{*}/\delta_{e}}\,
\sinh\left(\frac{x_{*}-x}{\delta}\right)}{\sinh
\left(\frac{x_{*}}{\delta}\right)}-e^{(x-x_{*})/\delta_{e}}\right]. \label{55}
\end{eqnarray}
If we let $x_{*}/\delta\to +\infty$, this formula becomes
\begin{eqnarray}
C =c_*\mu^2 {e^{-\xi/\mu}-e^{-\xi}\over\mu^2-1} = c_*\delta^2 {e^{(x-x_{*})
/\delta}- e^{(x-x_{*})/\delta_{e}}\over\delta^2-\delta^2_{e}}, \label{55bis}
\end{eqnarray}
which satisfies the conditions $C(\xi=0)=0=C(\xi=+\infty)$.

Now we need to calculate $x_{*}$ in such a way that the vapour flux $c'(x)$ is continuous at
$x=x_{*}$. Eq.\ (\ref{46}) indicates that $C$ should include a term of order $\epsilon$. 
The result is $(1-\epsilon T_{*})$ times $C$ in Eq.\ (\ref{55}). Inserting this into 
Eq.\ (\ref{46}), we obtain
\begin{eqnarray}
c'_{*} = {(1-\epsilon T_{*})\, c_{*}\over\delta + \delta_{e}},   \label{56}
\end{eqnarray}
where $\delta_{e}$ and $\delta$ are given by Eqs.\ (\ref{26bis}) and (\ref{43}), 
respectively. This is a modified version of the dew point shift equation (\ref{30}) to which 
it reduces if $\delta\ll\delta_{e}$. Eq.\ (\ref{56}) also holds for other flows if we 
interpret $c'_{*}$ as the normal derivative of the vapour density at the dew point interface. 

%\subsection{Perturbed dew point location}
To get $x_{*}$ in the case of a Hiemenz stagnation point flow, we solve  
\begin{eqnarray}
&& c'' + \mbox{Sc}\, u\, c'=0,  \quad\mbox{for}\quad x>x_{*},  \nonumber\\
&& c_{*}= {T_{d}\over T_{*}}\,\exp\left[{1\over\epsilon}\left({1\over T_{d}}
- {1\over T_{*}}\right)\right],\quad  c'_{*}= {(1-\epsilon T_{*})\, c_{*}\over 
{\epsilon T_{*}^2\over T'_{*}} + {1\over\sqrt{R\rho_*}}}, \quad\mbox{at $x=x_{*}$,}  
\label{57}
\end{eqnarray} 
instead of Eqs.\ (\ref{22}) and (\ref{30}). Then we calculate $c(+\infty)$, which is a 
function of $x_{*}$, and adjust $x_{*}$ until we obtain $c(+\infty)=1$. Once we know 
$x_{*}$, the vapour density profile in the CL is found using (\ref{55}):
\begin{eqnarray}
c(x)=c_{e}(x)+ \frac{c_*\delta^2(1-\epsilon T_{*})}{\delta^2-\delta^2_{e}}
\left[\frac{\sinh\left(\frac{x}{\delta}\right)+ e^{-x_{*}/\delta_{e}}\,
\sinh\left(\frac{x_{*}-x}{\delta}\right)}{\sinh
\left(\frac{x_{*}}{\delta}\right)}-e^{(x-x_{*})/\delta_{e}}\right]. \label{58}
\end{eqnarray}

\begin{figure}
\begin{center}
\includegraphics[width=10cm]{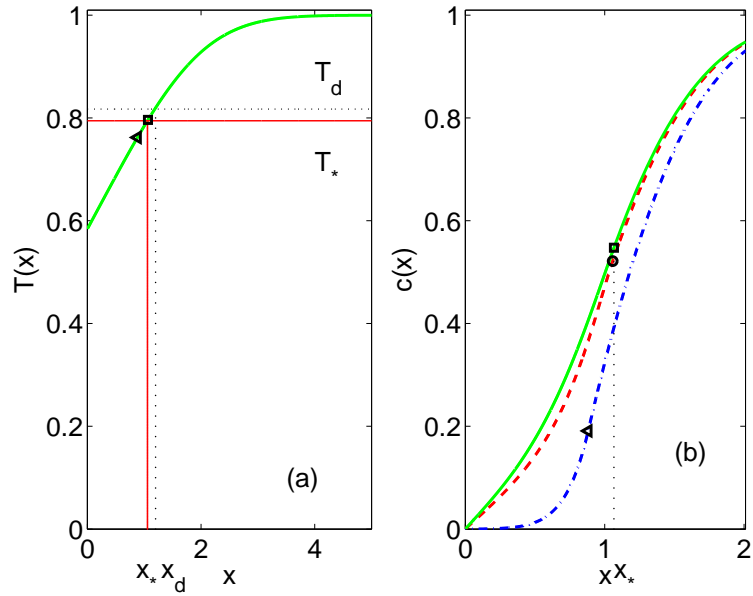}
\caption{Same as Figure \ref{fig1} for entry A in table 2. In (a), $T_{*}$ and $x_{*}$
are the values given by the $\delta$-CL theory, the square marks the value obtained by
numerically solving the complete model and the triangle marks the value provided by the 
0-CL theory. In (b), the thick solid line 
corresponds to the numerical calculation, the dashed line corresponds to Eq.\ (\ref{58}) 
($\delta$-CL) and the dot-dashed line to the 0-CL simplified equilibrium model. $x_{*}$ 
is marked by a square (numerical calculation of the full model), a circle ($\delta$-CL) and 
a triangle (0-CL). }
\label{fig2}
\end{center}
\end{figure}

These expressions give the decay of $c$ to the equilibrium density, $c_{e}(x)$. The function 
$C=c-c_{e}$ has a maximum at $x_{*}-[\delta\delta_{e}/(\delta-\delta_{e})]\ln(\delta/
\delta_{e})$, decays to zero as $(x-x_{*})\to -\infty$, and $c-c_{*}\sim (x_{*}-x)/
\delta$ as $x\to x_{*}-$.  As shown in figure \ref{fig2}, Eqs.\ (\ref{56}) and (\ref{58})
improve the approximation of $x_{*}$ and $T_{*}$ with respect to the simplified 
equilibrium model. 

\subsection{Deposition at the wall}
The vapour deposition rate and the total deposition rate at the wall for $T_{w}<T_{w,M}$ 
are given by inserting $c(x)$ given by (\ref{58}) in the exact equations (\ref{76bis}) and 
(\ref{77bis}), respectively. Then $J_{c}=J-J_{v}$. For $T_{w,M}\leq T_{w}<T_{d}$, 
the deposition rate is given by $J=[1-c_{e}(T_{w})]\, |\psi'(0)|/$Sc ($J_{c}=0$), in which
we again use $c_{e}(0)=c_{e}(T(0))=c_{e}(T_{w})$. Using (\ref{58}), we have 
observed that $J_{c}$ becomes zero for a certain $T_{w}$ (which gives the approximate
$T_{w,M}$ according to the $\delta$-CL theory). However, for this wall temperature
 $x_{*}>0$, and $J_{c}$ becomes negative for larger wall temperatures. Thus the $\delta$-CL
 theory gives unphysical results for the deposition rates for wall temperatures close to the
 wall temperature for which the numerical solution of the complete model yields $J_{c}=0$.

\section{Condensation layer for large $R$}
\label{sec:attached}
In the limit as $R\to +\infty$, the 0-CL theory gives an accurate description of the 
condensation layer. How do we correct this theory for large finite $R$?

Our idea is to use the method of {\em nonlinear} multiple scales in the limit as $R\to
+\infty$. The profiles $T$, $u$ and $\rho$ vary on the slow scale $x$ and we define
a fast nonlinear scale
\begin{eqnarray}
X= R^{1/2}\, g(x;1/R), \quad {dX\over dx}= R^{1/2}\, g', \label{62}
\end{eqnarray}
to be selected in such a way that the equation for the vapour concentration have constant
coefficients in the fast scale. The vapour density and $n$ are given by the expansions
\begin{eqnarray}
&& c= c_{e}+ R^{-1}\, C^{(0)}(X,x) + o(R^{-1}), \label{63}\\
&& n= 1+ O(R^{-1}). \label{64}
\end{eqnarray}
The choices $c-c_{e}=O(R^{-1})$ and $n-1=O(R^{-1})$ are dictated by dominant balance 
provided $N=O(1)$ as $R\to\infty$. Inserting Eqs.\ (\ref{63}) and (\ref{64}) in 
Eqs.\ (\ref{20}) and (\ref{21}), we find the following equations and boundary conditions
\begin{eqnarray}
&& g'^2\, \partial_{X}^2C^{(0)} - \rho\, C^{(0)} = - (c''_{e} + \mbox{Sc}\, u\, 
c'_{e}),     \label{65}\\
&& C^{(0)}(0,x)=0 =C^{(0)}(X_{*},x), \quad c(x_{*})= c_{e}(x_{*}).\label{66}
\end{eqnarray}
Let us select $g'=\sqrt{\rho}$, and therefore
\begin{eqnarray}
X=\int_{0}^x \sqrt{R\,\rho(x)}\, dx,\quad X_{*}= \int_{0}^{x_{*}} \sqrt{R\,
\rho(x)}\, dx, \label{71}
\end{eqnarray}
according to Eq.\ (\ref{62}). Since $\delta=(R\rho_{*})^{-1/2}\ll 1$ is a dimensionless decay
length for $C=c-c_{e}$ to vanish, $dX=dx\sqrt{R\rho}$ is a fast scale based on a space
dependent decay length. Using Eq.\ (\ref{71}), Eq.\ (\ref{65}) can be written as
\begin{eqnarray}
\partial_{X}^2C^{(0)} - C^{(0)} = - {c''_{e} + \mbox{Sc}\, u\, 
c'_{e}\over \rho},     \label{67}
\end{eqnarray}
whose left hand side has constant coefficients. The solution of Eqs.\ (\ref{67}) and (\ref{66}) is
\begin{eqnarray}
C^{(0)} = {c''_{e} + \mbox{Sc}\, u\, c'_{e}\over \rho} \left(1-{\sinh X +
\sinh(X_{*}-X)\over\sinh X_{*}}\right).    \label{68}
\end{eqnarray}
In principle, we should add a function of $x$ to the right hand side of Eq.\ (\ref{68}). However
the boundary conditions (\ref{66}) imply that such a function is identically zero. To find 
$n$, we integrate Eq.\ (\ref{17}) using the boundary condition (\ref{18}), $n(x_{*})=1$, 
and insert Eq.\ (\ref{68}) into the result, thereby obtaining
\begin{eqnarray}
n^{2/3} = 1 + {2N\over 3R}\int_{x}^{x_{*}}{c''_{e} + \mbox{Sc}\, u\, c'_{e}\over 
\rho U}\left(1-{\sinh X +\sinh(X_{*}-X)\over\sinh X_{*}}\right) dx + o(R^{-1}), 
\label{69}
\end{eqnarray}
and equivalently,
\begin{eqnarray}
n(x) = 1 + {N\over R}\int_{x}^{x_{*}}{c''_{e} + \mbox{Sc}\, u\, c'_{e}\over 
\rho U}\left(1-{\sinh X +\sinh(X_{*}-X)\over\sinh X_{*}}\right) dx + o(R^{-1}). 
\label{70}
\end{eqnarray}
Note that Eq.\ (\ref{70}) is consistent with Eq.\ (\ref{64}). 

\subsection{Vapour density profile and dew point location}
The vapour density profile is found from (\ref{68}) and (\ref{71}) as
\begin{eqnarray}
c(x) \sim c_{e}(x)+{c''_{e}(x) + \mbox{Sc}\, u(x) c'_{e}(x)\over R\rho(x)} \left(1-
{\sinh X(x) +\sinh[X_{*}-X(x)]\over\sinh X_{*}}\right).   
 \label{68bis}
\end{eqnarray}
The location $x_{*}$ of the dew point interface is found by imposing that the derivative of 
the vapour density be continuous there. To order $R^{-1/2}$, we have from Eq.\ 
(\ref{68bis}):
\begin{eqnarray}
c'_{*}= c'_{e}(x_{*}) -{c''_{e}(x_{*}) + \mbox{Sc}\, u_{*} c'_{e}(x_{*})\over 
\sqrt{R\rho_{*}}}\, {\cosh X_{*}-1\over\sinh X_{*}}.     \label{72}
\end{eqnarray}
We have omitted the terms of order $1/R$ because (\ref{68bis}) does not include
corrections of order $R^{-3/2}$, and such corrections also contribute $O(1/R)$ terms to 
$c'(x)$. Similarly, at the wall we have
\begin{eqnarray}
c'(0)= c'_{e}(0) + {c''_{e}(0)\over\sqrt{R\rho(0)}}\, {\cosh X_{*}-1\over\sinh 
X_{*}}.     \label{73}
\end{eqnarray}
To get $x_{*}$ in the case of a Hiemenz stagnation point flow, we solve Eq.\ (\ref{22}) with
$c_{*}=c_{e}(x_{*})$ and (\ref{72}) with $X_{*}$ given by Eq.\ (\ref{71}). 

\subsection{Deposition at the wall}
Using Eq.\ (\ref{76bis}) and (\ref{73}), we calculate $J_{v}$ and using Eq.\ 
(\ref{70}) in Eq.\ (\ref{76}), we find $J_{c}$. We obtain 
\begin{eqnarray}
J_{v}&=& \frac{1}{\mbox{Sc}}\left(c'_{e}(0) + {c''_{e}(0)\over\sqrt{R\rho(0)}}\,
{\cosh X_{*}-1\over\sinh X_{*}}\right) + O\left(\frac{1}{R}\right), 
\label{80a}\\
J_{c}&=& {\rho(0)U(0)\over\mbox{Sc}}\int_{0}^{x_{*}}{c''_{e} + \mbox{Sc} u 
c'_{e}\over\rho U}\left(1-{\sinh X +\sinh(X_{*}-X)\over\sinh X_{*}}\right) dx
+O\left(\frac{1}{\sqrt{R}}\right),\quad  \label{80}
%J&=& \frac{\rho(0)U(0)}{\mbox{Sc}}\left\{\frac{c'_{*}}{\rho_{*}U_{*}} +
%\int_{0}^{x_{*}}\frac{\mbox{Sc}\, uU+u'}{\rho U^2}\left[c'_{e}+ \frac{
%\cosh(X_{*}-X)-\cosh X}{\sinh X_{*}} \right.\right.\\
%&\times& \frac{c''_{e}+\mbox{Sc}\, uc'_{e}}{\sqrt{R\rho}}
%+ \left.\left. \left(\frac{c''_{e}+\mbox{Sc}\, uc'_{e}}{R\rho}\right)'
%\left(1-{\sinh X +\sinh(X_{*}-X)\over\sinh X_{*}}\right) \right]dx\right\}. \label{80b}
\end{eqnarray}
and $J=J_{v}+J_{c}$. As $R\to +\infty$, Eqs.\ (\ref{80a}) and (\ref{80}) coincide
with the corresponding expressions of the 0-CL theory. 
 
\section{Numerical results}
\label{sec:numerical}
In this Section we shall compare the location of the dew point interface, the dew point 
temperature shift and the deposition flux at the wall obtained by the asymptotic theories
of Sections \ref{sec:equilibrium}, \ref{sec:condlayer} and \ref{sec:attached} to the 
values given by a direct numerical solution of the free boundary problem (\ref{15}) - 
(\ref{23}) for stagnation point flow. We have considered four representative parameter 
choices to illustrate the ranges of validity of our different asymptotic approximations. 

Firstly in table 3, we use $\tilde{T}_{\infty}=1713$ K, $\tilde{T}_{d}=1400$ K and 
$\tilde{T}_{w}=1000$ K for two choices of $R$. Choice A has $R=4.93$ leading to 
relatively large width of the condensation layer (in which $c\neq c_{e}$), whereas $R$ is 150 
times larger for Choice C leading to a very narrow condensation 
layer. In both cases, the CL is relatively detached from the wall ($x_{*}/\delta_{e}\approx 
5.87$ and 5.67 for cases A and C, respectively), but setting $x_{*}/\delta_{e}=+
\infty$ as in Eq.\ (\ref{55bis}) yields poor approximations for the vapour profile and
the deposition rates. Table 3 compares the results given by the simplified 
equilibrium theory (0-CL), the $\delta$-CL theory given by Eq.\ (\ref{58}) and related 
ones, by the nonlinear multiple scales theory (NLMS, choice C only) and by 
direct numerical simulation of the problem (\ref{15}) - (\ref{23}). As $R$ increases,
the shooting problem which yields $x_{*}$ is ill conditioned. Then we need to calculate many
significant digits of $x_{*}$ to get a good approximation of the deposition rates $J_{v}$,
$J_{c}$ and $J$ in table 3. The $x_{*}$ values are given with four digits in table 3, but
we have calculated them with six and twelve digits for parameter choices A and C, 
respectively.
\begin{table}%[h]
\begin{center}
\begin{tabular}{@{}c|cccccl@{}}    %{|c|c|c|}
%\hline
-- & {0-CL} &{$\delta$-CL (A)}& {num.\ (A)} & {$\delta$-CL (C)} & {NLMS (C)}
& {num.\ (C)}\\[3pt]
%\hline
$x_{*}$ & 0.8815 &1.0574 &1.0666 & 0.9152&0.9212 & 0.9146\\
%\hline
$T_{*}$ & 0.7620 &0.7946 & 0.7962&0.7683 & 0.7695&0.7682\\
%\hline
$T'_{*}$ & 0.1902 &0.1803 & 0.1797 &0.1885& 0.1882 &0.1886\\
%\hline
$u_{*}$& 0.3700 &0.5045 & 0.5119&0.3948 & 0.3992&0.3943\\
%\hline 
$U_{*}$ &0.3950 &0.5272& 0.5344&0.4193&0.4237 &0.4189 \\
%\hline 
$c_{*}$ & 0.1910 &0.5215& 0.5475&0.2341 &0.2426&0.2332\\
%\hline 
$c'_{*}$& 1.1676&0.7875 & 0.7499 &1.1327&1.1240 & 1.1334 \\
 %\hline
 $\rho_{*}$ &0.9758&0.9807& 0.9809&0.9768 &0.9770& 0.9768\\       
%\hline
$\delta$ & --  & 0.4548 & 0.4547 &0.0372 &0.0372&0.0372\\       
%\hline
$\delta_{e}$&0.1572&0.1803&0.1817&0.1613 &0.1620&0.1612 \\       
%\hline
$J_{v}$ & 0.0007 &0.1365& 0.1658& 0.0015& 0.0009& 0.0045\\       
%\hline
$J_{c}$ &0.0876  &0.0694& 0.0618&0.0879& 0.0886 &0.0849\\       
%\hline
$J$ &0.0883  &0.2059& 0.2276&0.0894 & 0.0896 &0.0894\\       
%\hline
\end{tabular}
\label{results}
\caption{Dimensionless results for wall temperature $T_{w}=0.5838$ (1000 K). Results 
obtained with the simplified equilibrium theory corresponding to a zero-width CL are in the
column 0-CL, those obtained with matched asymptotic expansions for a CL of width $\delta$
are in the column $\delta$-CL, whereas NLMS refers to corrections to the equilibrium
theory obtained using the method of nonlinear multiple scales. Results obtained by solving
numerically the complete model are indicated by ``num.''. In this table, $(R,N)$ are 
(4.93,0.11) and (739.5,1.65) for entries labeled A and C respectively. The values of 
$\alpha$, $\epsilon$, Pr and Sc are as in table 2.}
\end{center}
\end{table}

For low wall temperature and small values of $R$, the 0-CL theory gives
much worse values of $x_{*}$ than the $\delta$-CL theory. This is also shown in
figure \ref{fig2}: the 0-CL theory yields a vapour density curve below the others. Eq.\
(\ref{58}) provides the best approximation to the numerical solution of the
complete problem. According to our expectations, the simplified equilibrium theory is a good
approximation for large values of $R$, cf.\ figure \ref{fig1}. Table 3 shows that the
three asymptotic theories, 0-CL, $\delta$-CL and NLMS, underestimate the flux $J_{v}$
and overestimate $J_{c}$, thereby yielding reasonable values of the total deposition rate $J$.
For this low $T_{w}$, the $\delta$-CL theory gives the best results and the 0-CL theory
the worse ones. 

\begin{table}%[h]
\begin{center}
\begin{tabular}{@{}c|cccccl@{}}    %{|c|c|c|}
%\hline
-- & {0-CL} &{$\delta$-CL (D)}& {num.\ (D)}&{$\delta$-CL (E)} & {NLMS (E)}
& {num.\ (E)}\\[3pt]
%\hline
$x_{*}$ & 0.3900&0.6074 &0.5694 &0.4283 & 0.4305& 0.4252\\
%\hline
$T_{*}$ & 0.7579&0.7897 &0.7842 &0.7635& 0.7639 &0.7631\\
%\hline
$T'_{*}$ & 0.1476 &0.1446&0.1453 & 0.1472&0.1472& 0.1473\\
%\hline
$u_{*}$& 0.0840&0.1909 &0.1697&0.1001 &0.1011 &0.0988\\
%\hline 
$U_{*}$ &0.1034 &0.2092 &0.1882&0.1194&0.1204 & 0.1180\\
%\hline 
$c_{*}$ & 0.1675&0.4512&0.3823&0.2010& 0.2031& 0.1980\\
%\hline 
$c'_{*}$&0.8033&0.6390 &0.6935&0.8004 &0.7980 &0.8000\\
% \hline
 $\rho_{*}$ & 0.9708 &0.9784&0.9773&0.9724&0.9725 &0.9723\\       
%\hline
$\delta$ & -- &0.4553 &0.4556 &0.0373 &0.0373 & 0.0373\\       
%\hline
$\delta_{e}$& 0.2004 &0.2221 &0.2180 &0.2041&0.2039 & 0.2036\\       
%\hline
$J_{v}$ & 0.0688&0.3172 &0.3148&0.0832&0.0826 &0.0919\\       
%\hline
$J_{c}$ & 0.1627&0.0492&0.0316&0.1524&0.1486 & 0.1437\\       
%\hline
$J$ & 0.2315&0.3664&0.3464&0.2356&0.2312 & 0.2356\\       
%\hline
\end{tabular}
\label{results2}
\caption{Dimensionless results for wall temperature $T_{w}=0.7006$ (1200 K). Case D
corresponds to $R=4.93$ and $N=0.11$ whereas Case E corresponds to $R=739.5$ and
$N=1.65$. The values of $\alpha$, $\epsilon$, Pr and Sc are as in table 2.}
\end{center}
\end{table}

\begin{figure}
\begin{center}
\includegraphics[width=10cm]{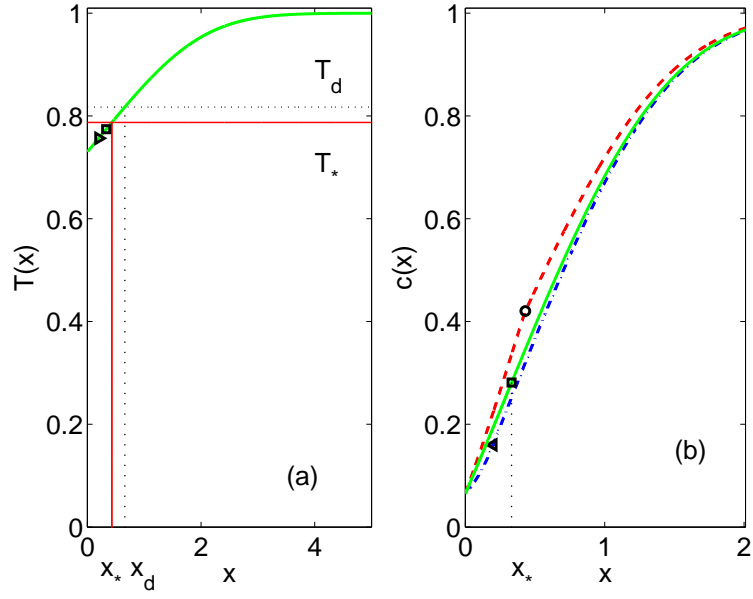}
\caption{Same as Figure \ref{fig2} for entry A in table 2 and $T_{w}=0.7298$ (1250 K).
In (b) we have depicted the numerical solution of the full model (solid line, square), Eqs.\
(\ref{57}) - (\ref{58}) (dashed line, circle) and the 
simplified equilibrium model (dot-dashed line, triangle). }
\label{fig3}
\end{center}
\end{figure}

\begin{figure}
\begin{center}
\includegraphics[width=10cm]{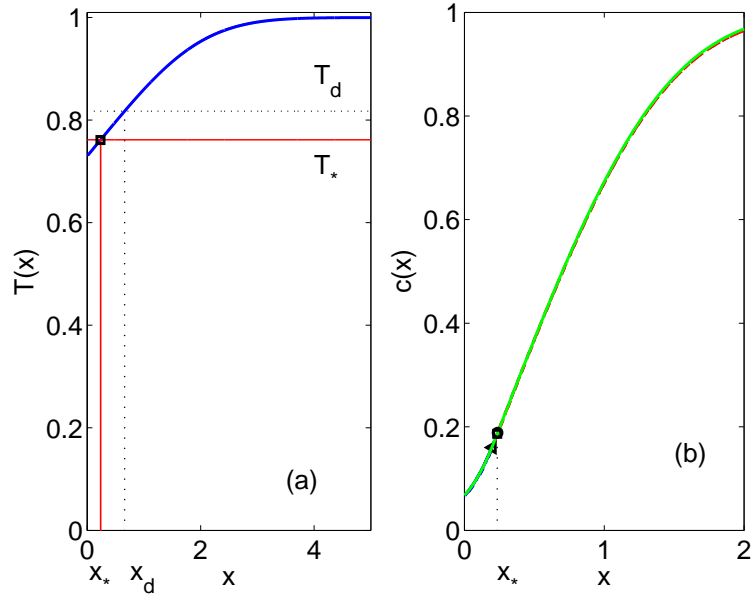}
\caption{Same as Figure \ref{fig2} for entry C in table 2 and $T_{w}=0.7298$. In (b), 
the vapour density profiles provided by the 0-CL, $\delta$-CL and NLMS theories overlap 
the numerical solution of the full model. However, $x_{*}$ as
calculated using the 0-CL theory does not approximates the numerical value as
precisely as the calculation using the $\delta$-CL and NLMS theories. }
\label{fig4}
\end{center}
\end{figure}

Figures \ref{fig3} and \ref{fig4} depict the vapour density for $R$ and $N$ as in entries A 
and C of table 3 but for a higher $T_{w}=0.7298$ ($\tilde{T}_{w}=$ 1250 K) close to 
$T_{w,M}=0.755$. The dew point interface is closer to the wall ($x_{*}/\delta_{e}
\approx 1.43$ 
and 2.09 for cases A and C, respectively). For parameter choice A, $c(x)$ is approximated
better by the 0-CL theory than by the $\delta$-CL theory, whereas for the larger $R$ of 
choice C, all three asymptotic theories approximate well the numerical result. The poorer 
performance of the $\delta$-CL theory for parameter choice A is a consequence of the
fact that $T_{w}$ is close to $T_{w,M}$. For somewhat lower $T_{w}=0.7006$ (1200 K),
$x_{*}/\delta_{e}=2.61$ (choice A) and 2.09 (choice C). Table 4 shows that the 
$\delta$-CL theory gives the best approximation to the deposition rates. As before in table 3, 
the $x_{*}$ values are given with four digits in table 4, but we have calculated them with 
five and seven digits for parameter choices D and E, respectively. For the higher $T_{w}$
in table 4, $x_{*}$ is smaller than in table 3 and less precision is needed to calculate the
deposition rates.

%According to table 4, the dew point interface comes closer to the wall while 
%the decay lengths remain comparable to $x_{*}$ ($\delta_{e}\approx x_{*}<\delta$). 
%Then the wall is met before $c$ has had time to decay to $c_{e}$ according to the theory in 
%Section \ref{sec:condlayer}. In particular, the flow velocity and $U(x)$ cannot be taken as 
%constants in the condensation layer as assumed there, and therefore the expressions given by 
%this theory are now poor approximations. Figure \ref{fig3} shows that the simplified 
%equilibrium theory gives a good approximation to the vapour density. The reason is that $c=
%c_{e}$ both at $x_{*}$ and at the wall, so that $c\approx c_{e}$ between $x=0$ and 
%$x_{*}$. Figure \ref{fig4} shows that all our asymptotic theories well approximate 
%the vapour density for large $R$ as expected, given the very small value of the decay length 
%$\delta$. For large $R$ (equivalently, $\delta\to 0$), using nonlinear multiple scales as in 
%Section \ref{sec:attached} gives the best approximation for a condensation layer extending 
%to the wall. Table 3 shows that the theory of nonlinear multiple scales gives the 
%most reliable approximation to the fluxes $J_{v}$ and $J_{c}$ of all our theories in the limit 
%of large $R$.

\begin{figure}
\begin{center}
\includegraphics[width=10cm]{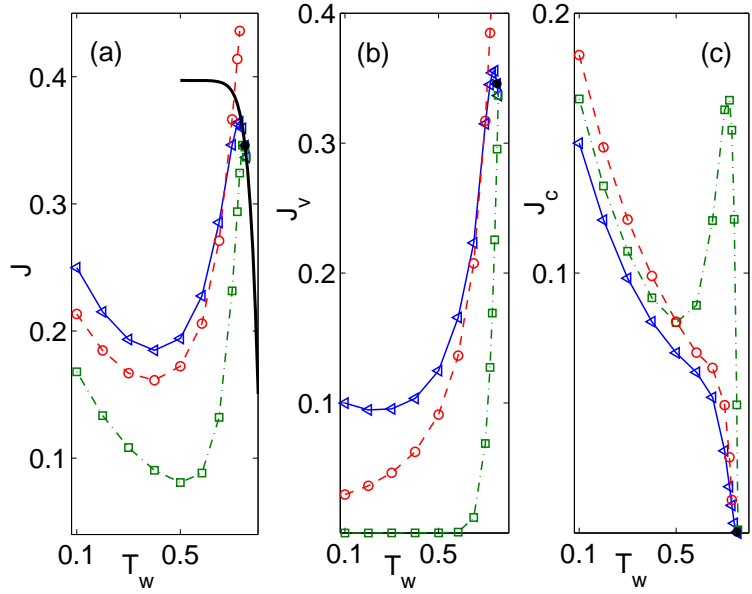}
\caption{Deposition rates $J$, $J_{v}$ and $J_{c}$ as a function of the wall temperature
for $R=4.93$, $N=0.11$. Solid line: numerical solution of the full model (triangles), dashed 
line: $\delta$-CL theory (circles), dot-dash line: equilibrium (squares). The thick solid
line are the exact deposition rate calculated with $J_{c}=0$. The NLMS approximation can 
only be calculated for the wall temperature marked with an asterisk and higher ones. The 
values of $\alpha$, $\epsilon$, Pr and Sc are as in table 2. }
\label{fig5}
\end{center}
\end{figure}

\begin{figure}
\begin{center}
\includegraphics[width=10cm]{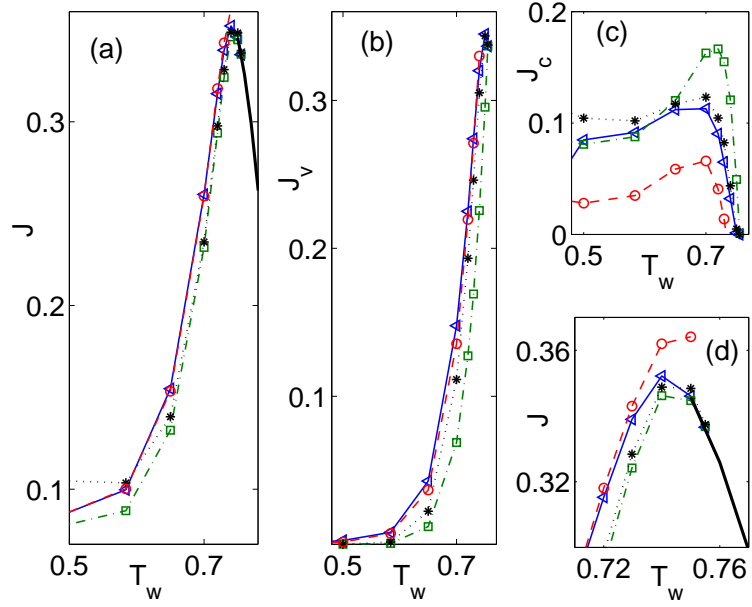}
\caption{ Deposition rates (a) $J$, (b) $J_{v}$ and (c) $J_{c}$ as functions of the wall 
temperature for $R=73.95$, $N=0.11$. (d) is a zoom of (a) near the maximum value of $J$. 
Solid line: numerical solution of the full model (triangles), dashed line: $\delta$-CL theory 
(circles), dot-dash line: 0-CL (squares). The thick solid lines in 
(a) and (d) are the exact deposition rates for $T_{w,M}\leq T_{w}<T_{d}$ where $J_{c}=
0$. The values of $\alpha$, $\epsilon$, Pr and Sc are as in table 2. }
\label{fig6}
\end{center}
\end{figure}

\begin{figure}
\begin{center}
\includegraphics[width=10cm]{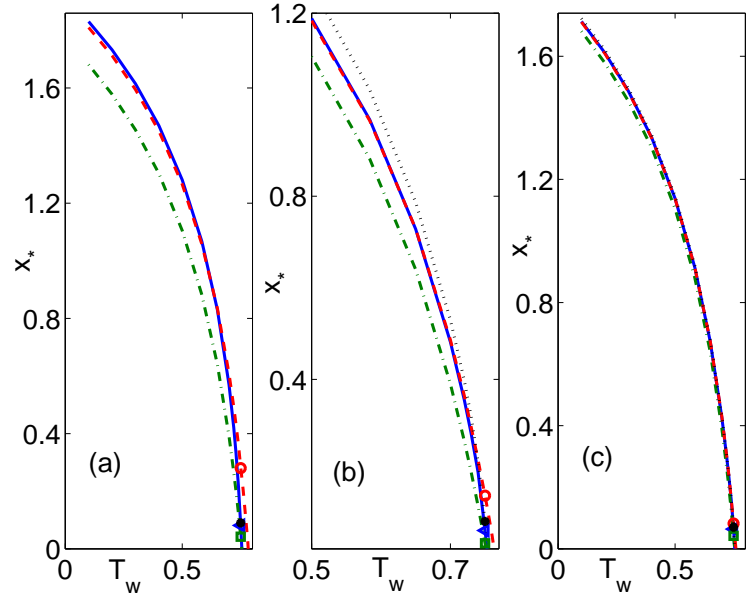}
\caption{Dew point location vs wall temperature for (a) $R=4.93$, $N=0.11$, (b) $R=73.95$,
$N=0.11$, and (c) $R=739.5$, $N=1.65$. Values of $\alpha$, $\epsilon$, Pr and Sc are as in table 2. 
Data are obtained from the numerical solution of the complete model (solid line, triangle), the 
0-CL theory (dot-dash line, square), the $\delta$-CL theory (dash line, a circle) and NLMS 
(dot line, asterisk). The triangle, square, circle and asterisk correspond to the same wall
temperature.}
\label{fig7}
\end{center}
\end{figure}

\begin{figure}
\begin{center}
\includegraphics[width=10cm]{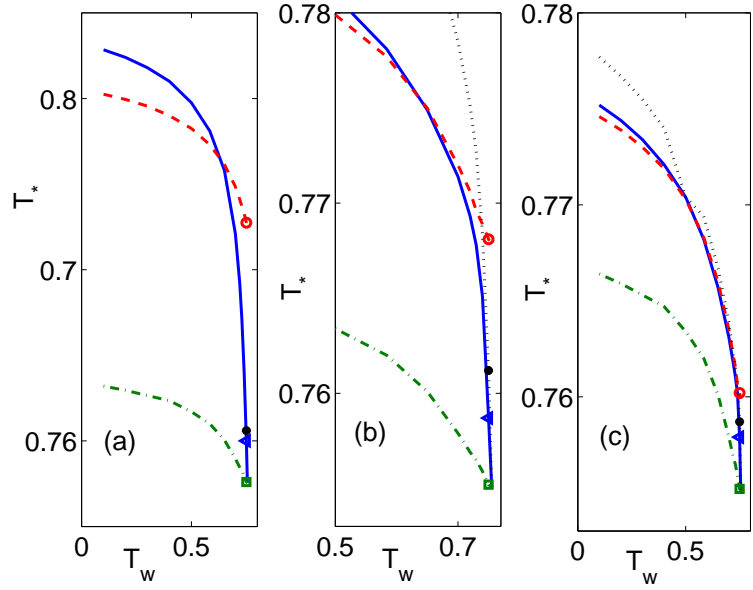}
\caption{Same as Figure \ref{fig7} for the dew point temperature vs wall temperature
diagram.}
\label{fig8}
\end{center}
\end{figure}

\begin{figure}
\begin{center}
\includegraphics[width=10cm]{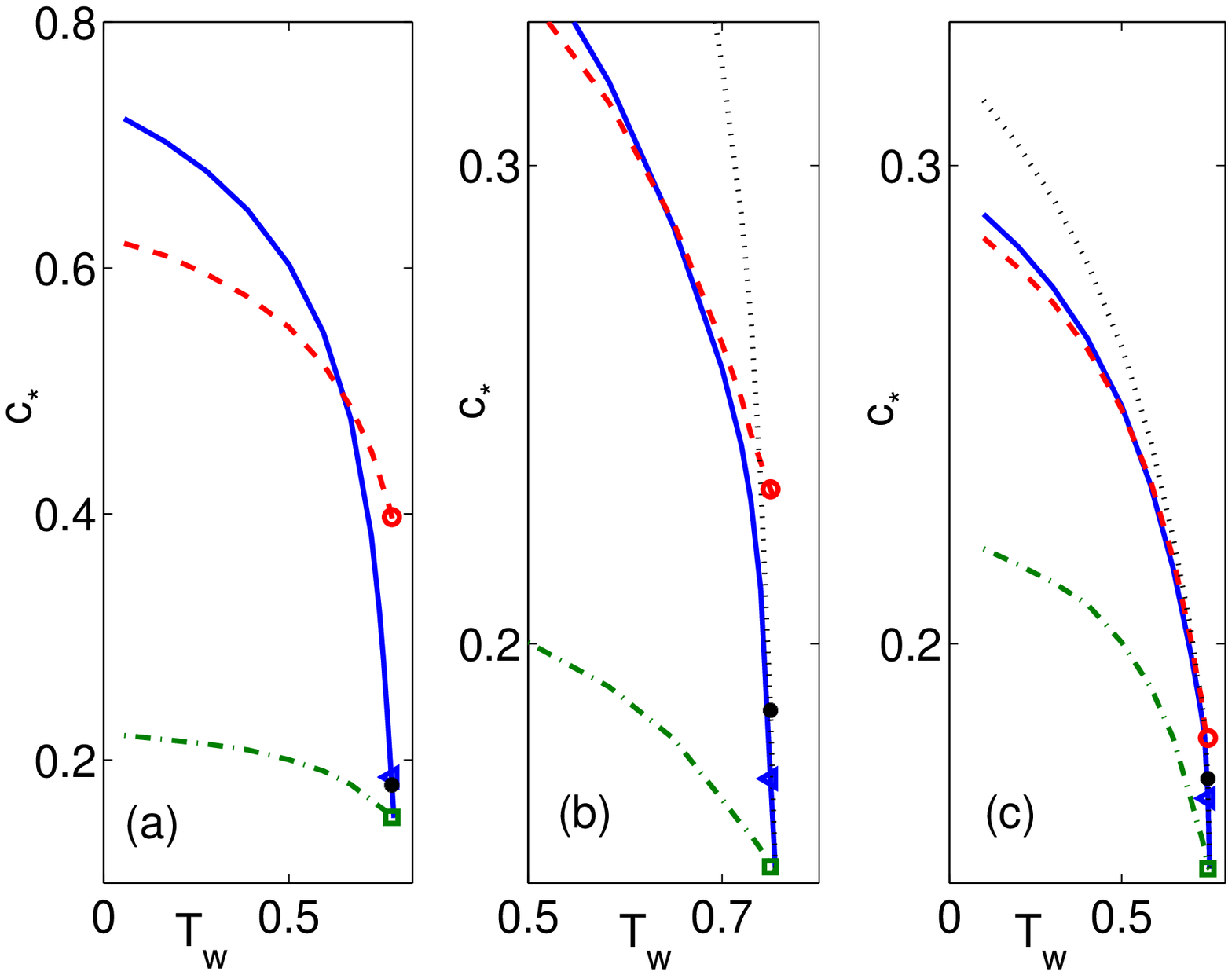}
\caption{Same as Figure \ref{fig7} for the vapour concentration at $x_{*}$ vs wall 
temperature diagram.  }
\label{fig9}
\end{center}
\end{figure}

Figures \ref{fig5} to \ref{fig9} show the dependence of deposition rates, dew point
location, $T_{*}$ and $c_{*}$ with $T_{w}$. They can be used to compare the different
asymptotic theories. Figures \ref{fig5}(a) and \ref{fig6}(a) show that there is a value
of $T_{w,M}$ for which $J_{c}=0$: at this value, the thick solid line representing $J_{c}=0$,
Eq.\ (\ref{78}), departs from the numerical $J$. This value is overestimated by the 0-CL
theory and underestimated by the $\delta$-CL theory\footnote{The $\delta$-CL theory 
predicts $J_{c}=0$ with $x_{*}\neq 0$ for a wall temperature smaller than the numerical
$T_{w,M}$. For larger $T_{w}$, the $\delta$-CL theory yields unphysical rates $J_c<0$ 
and smaller $x_{*}$ which eventually become zero.}, whereas the NLMS theory gives a 
somewhat better prediction. For the parameter choice A in table 3, Fig.\ \ref{fig5} indicates 
that the $\delta$-CL theory is better than the 0-CL theory for $T_{w}<T_{w,M}$ but 
not very close to $T_{w,M}$. Fortunately, in a small neighborhood of $T_{w}=T_{w,M}$ 
the 0-CL theory and its correction by NLMS provide a good approximation to the deposition 
rates. For $T_{w}>T_{w,M}$, Eq.\ (\ref{78}), or equivalently $J=[1-c_{e}(T_{w})]\, 
|\psi'(0)|/$Sc (obtained by setting $J_{c}=0$), is exact. As we may have expected, all three 
asymptotic theories provide good approximations to $J_{v}$, $J_{c}$ and $J$ for large $R$, 
as shown in Fig.\ \ref{fig6}. 

In Figure \ref{fig7}, we show the dew point location in terms of $T_{w}$. For low $R$, the 
$\delta$-CL theory approximates better the numerical $x_{*}$ except very close to its 
estimated value of $T_{w,M}$ (for which the $\delta$-CL theory gives $J_{c}=0$ with 
$x_{*}\neq 0$, marked with a circle in the figure). But for such values of $T_{w}$, the 
NLMS theory takes over and it yields a good approximation to the numerical value of 
$x_{*}$. Figures \ref{fig7}(b) and (c) show that the NLMS theory is a good approximation 
to the numerical $x_{*}$ as $T_{w}$ approaches $T_{w,M}$ for intermediate and for large 
values of $R$. Similarly, Figures \ref{fig8} and \ref{fig9} show that the NLMS 
approximates well $T_{*}$ and $c_{*}$ for $T_{w}$ very close to $T_{w,M}$\footnote{
This is so also for small $R$ in Figs.\ \ref{fig7}(a), \ref{fig8}(a) and \ref{fig9}(a), in 
which the NLMS approximation can be calculated only for sufficiently large wall 
temperatures, close to $T_{w,M}$. Below a certain wall temperature, Eq.\ (\ref{72}) does 
not have a solution, and therefore the NLMS theory does not provide an approximate 
$x_{*}$.}. At a lower wall temperature, $T_{w,c}$, the values of $c_{*}$ and $T_{*}$ 
given by the $\delta$-CL and NLMS theories coincide. A good compromise to attain a 
uniform approximation could be to use the $\delta$-CL theory for $T_{w}<T_{w,c}$, the 
NLMS theory for $T_{w,c}<T_{w}<T_{w,M}$ and (\ref{78}) for $T_{w,M}<T_{w}<
T_{d}$. 

\section{Discussion}
\label{sec:final}
We have considered heterogeneous condensation of vapours mixed with a carrier gas in the stagnation
point boundary layer flow near a cold wall. For the case of Na$_{2}$SO$_{4}$ vapours in air with
a diluted suspension of solid particles with radius one micron, the mean free path of vapour particles is
one tenth of the particle radius, so we have assumed that the supersaturated vapour condenses on the particles
by diffusion. This is different from the kinetic theory formulas used by \cite{CR88}
and later authors, \cite{fil03}, which are valid in the opposite limit in which the mean free path is much
larger than the particle size. The particles and droplets move towards the wall by thermophoresis and the 
Soret and Dufour effects have been ignored. We have assumed that the heat of vaporization 
is much larger than the Boltzmann constant times the temperature far from the wall. Under 
these conditions, vapour condensation occurs in a condensation layer whose distance to the wall, 
width and characteristics depend on the parameters of the problem. 

We have presented different asymptotic theories of the condensation process, calculated the 
shift in the dew point interface due to the flow, the vapour density profile and the deposition 
rates at the wall and compared them to direct numerical simulation of the equations governing the 
model. The simplest 0-CL theory, already studied by Castillo and Rosner, assumes that the 
width of the condensation layer is zero and that the vapour is in equilibrium with the 
condensed liquid in the dew surface. A more complete $\delta$-CL theory considers a 
condensation layer of finite width within which the vapour density has not yet reached local 
equilibrium with the liquid. In the CL, temperature, velocity and droplet density are 
approximated by their constant values at the dew point interface and the vapour density 
satisfies a linear equation. The deposition rates at the wall are calculated using this approximate
profile in exact expressions for the rates. The $\delta$-CL theory approximates well the
numerical vapour profile and deposition rates (better than the 0-CL theory) except in a narrow 
interval of wall temperatures near a maximum value $T_{w,M}$ at which the deposition rate 
$J_{c}$ of vapour coated droplets becomes zero. In the limit as the product $R$ of particle density, 
particle radius and the square of the Hiemenz width tends to infinity, the width of the CL tends 
to zero and an asymptotic calculation based on nonlinear multiple scales approximates well the 
vapour density and deposition fluxes given by a numerical solution of the full set of model 
equations. For moderate and low values of $R$, the multiple scales calculation holds for wall
temperatures close to the maximum one and corrects there the $\delta$-CL theory. If we
denote by $T_{w,c}$ the wall temperature at which the multiple scales and $\delta$-CL 
theories produce the same value of $x_{*}$, we obtain a uniform approximation
to the deposition rate by using the $\delta$-CL theory for $T_{w}<T_{w,c}$, the NLMS
theory for $T_{w,c}<T_{w}<T_{w,M}$ and the exact expression (\ref{78}) for the case
$J_{c}=0$ if $T_{w,M}<T_{w}<T_{d}$ ($T_{d}$ is the dew point temperature in the
absence of flow). Note that for large $R$, all three asymptotic theories yield reasonable
values of the deposition rates for almost any wall temperature because $T_{w,M}-T_{w,c}$
is very small and the correction to the 0-CL theory given by NLMS vanishes as $R\to\infty$.

A more complete thermophysical model of heterogeneous condensation and deposition of 
condensed vapour on cold walls is due to \cite{GR86} who 
assumed that the viscosity, thermal conductivity, specific heat and density of the carrier
gas depend on powers of the temperature $\tilde{T}$, and so does the diffusion coefficient 
of the vapour. In addition, the thermophoretic coefficient is $\alpha\propto 1-C/\tilde{T}$.
A simpler version of this model was used by \cite{fil03} to analyze the OVD 
process. This author considers that the carrier gas density may vary in the boundary layer and 
that its viscosity is proportional to $\tilde{T}^m$, where $m$ varies between 0.5 and 0.7. 
In addition, the flux of vapour includes thermal diffusion (Soret effect) and the rate of vapour 
condensation on a spherical solid particle is given by the kinetic theory formula in 
\cite{CR88}, assuming that the mean free path is much larger than particle size (different from the case we 
consider in the present paper). Of all these additional processes, considering a 
temperature-dependent viscosity produces the greatest effects in particle concentrations and 
deposition fluxes, \cite{fil03}. For the OVD process, the particle density $\tilde{\rho}_{
\infty}$ is much larger than the values we have considered here, which results in values of 
$R$ much larger than those considered in the present paper. Filippov's analysis uses a fast 
linear multiple scale $\xi=\sqrt{R}x$ (in our notation) instead of our scale $X=\int_{0}^x
\sqrt{R\rho}\, dx$ in the limit as $R\to +\infty$, $N=O(R)$, thereby obtaining a 
solution $C^{(0)}$ that contains exponentials of $x\sqrt{R\rho(x)}$ instead of $X$. If we 
use Filippov's multiple scales in our simpler thermophysical model, the equation for $C^{(0)}$ 
such that $c-c_{e}\sim R^{-1}C^{(0)}$ and $n-1\sim R^{-1}n^{(0)}$ becomes
\begin{eqnarray}
\partial_{\xi}^2C^{(0)} - \rho\, C^{(0)} = - (c''_{e} + \mbox{Sc}\, u\, c'_{e}).
\label{65f}
\end{eqnarray}
instead of (\ref{65}), and the boundary conditions for $C^{(0)}$ are still (\ref{66}). The
solution of this boundary value problem is 
\begin{eqnarray}
C^{(0)} = {c''_{e} + \mbox{Sc}\, u\, c'_{e}\over \rho} \left(1-{\sinh[\xi\sqrt{
\rho(x)}] + \sinh[(\xi_{*}-\xi)\sqrt{\rho(x)}]\over\sinh[\xi_{*}\sqrt{\rho(x)}]}
\right),    \label{68f}
\end{eqnarray}
instead of Eq.\ (\ref{68}). Eq.\ (\ref{68f}) corresponds exactly to Equation (70) for
$c-c_{e}$ in \cite{fil03}. Unless $\rho$ is constant (cf.\ $l(\eta)$ is constant in 
\cite{fil03}), Filippov's results are inconsistent: calculation of the next order correction 
$C^{(1)}$ in $c-c_{e}\sim R^{-1}C^{(0)}+ R^{-3/2}C^{(1)}$, which solves
\begin{eqnarray}
\partial_{\xi}^2C^{(1)} - \rho\, C^{(1)} = -2\partial_{\xi}\partial_{x}C^{(0)}
- \mbox{Sc}\, u\,\partial_{\xi}C^{(0)},  \label{66f}
\end{eqnarray}
 would give terms proportional to $\xi^2$. Then $R^{-3/2}C^{(1)}$ would contain terms 
 proportional to $R^{-1/2}x^2$ which are not small compared to $R^{-1}C^{(0)}$. 
 Inconsistency is thus tracked to the mixture of slow and fast scales in the solution $C^{(l)}$, 
$l=0,1$. The same mixture of scales also occurs for the droplet radius (equivalent to our 
$n^{1/3}$). In \cite{fil03}, the results of the analysis are not compared to a direct numerical 
solution of the complete thermophysical model. Then we do not know whether the 
perturbation method used in that paper produces results that at least give the correct order of 
magnitude. It is clear that the methods explained in the present work can be applied to the 
more detailed thermophysical model of \cite{fil03} or to that of \cite{GR86}. 

%{\bf ADDITIONAL STUFF FROM APPROXIMATING THE GREEN FUNCTION:} I get:
%\begin{eqnarray}
%C=c-c_{e} = \frac{1}{R}\left[{c''_{e} + \mbox{Sc}\, u\, c'_{e}\over \rho} -
%\frac{c''_{e}(0)\exp\left(-\frac{\mbox{Sc}}{2}\int_{0}^{x}u\, dx\right)}{
%\rho^{1/4}\rho(0)^{3/4}} {\sinh(X_{*}-X)\over\sinh X_{*}} \right. \nonumber\\
%\left. - {c''_{e}(x_{*}) + \mbox{Sc}\, u_{*} c'_{e}(x_{*})\over \rho_{*}^{3/4}
%\rho^{1/4}}\frac{\sinh X\,\exp\left(\frac{\mbox{Sc}}{2}\int_{x}^{x_{*}}u\, 
%dx\right)}{\sinh X_{*}}\right]+ O(R^{-3/2}),    \nonumber
%\end{eqnarray}
%instead of (\ref{68}). The fluxes are:
%\begin{eqnarray}
%J_{v}= \frac{c'_{e}(0)}{\mbox{Sc}}+\frac{c''_{e}(0)}{\mbox{Sc}\,\sqrt{R\rho(0)}}
%\coth X_{*}- \frac{c''_{e}(x_{*})+\mbox{Sc}\,u_{*}c'_{e}(x_{*})}{\mbox{Sc}\,
%\sqrt{R}\,\rho_{*}^{3/4}\sinh X_{*}} [\rho(0)]^{1/4}\exp\left(\frac{\mbox{Sc}}
%{2}\int_{0}^{x_{*}}u\, dx\right),  \nonumber\\
%J_{c}=\frac{\rho(0)U(0)}{\mbox{Sc}}\left\{\int_{0}^{x_{*}}\frac{c''_{e}+
%\mbox{Sc}\,u c'_{e}}{\rho U}\, dx - \left(\frac{c''_{e}(x_{*})+\mbox{Sc}\,u_{*}
%c'_{e}(x_{*})}{\rho_{*}^{3/2}U_{*}}+ \frac{c''_{e}(0)}{\rho(0)^{3/2}U(0)}\right)
%\frac{\coth X_{*}}{\sqrt{R}}\right.\nonumber\\
%+\frac{1}{\sqrt{R}[\rho_{*}\rho(0)]^{3/4}\sinh X_{*}}\left[\frac{c''_{e}(0)}{
%U_{*}} \exp\left(-\frac{\mbox{Sc}}{2}\int_{0}^{x_{*}}u\, dx\right) \right.\nonumber\\
%\left.\left. +\frac{c''_{e}(x_{*})+\mbox{Sc}\,u_{*}c'_{e}(x_{*})}{U(0)} 
%\exp\left(\frac{\mbox{Sc}}{2}\int_{0}^{x_{*}}u\, dx\right)\right]\right\}. \nonumber
%\end{eqnarray}
%up to terms of order $R^{-1}$.

\begin{acknowledgments}
We thank J.L. Castillo, Y. Farjoun and P.L. Garcia Ybarra for fruitful discussions and useful 
suggestions. We also thank the referees for useful comments and suggestions. This work has 
been supported by the National Science Foundation Grant DMS-0515616 (JCN), by the 
Ministry of Science and Innovation grants FIS2008-04921-C02-02 (AC) and 
FIS2008-04921-C02-01 (LLB), and by the Autonomous Region of Madrid Grant 
S-0505/ENE/0229 (COMLIMAMS) (LLB and JCN).
\end{acknowledgments}

\noindent {\Large\bf FIGURES}
\newpage


\begin{thebibliography}{13}
\bibitem[Batchelor \& Shen(1985)]{BS85}
\textsc{Batchelor, G. K. \& C. Shen, C.} 1985
Thermophoretic deposition of particles in gas flowing over cold surfaces,
\emph{J.~Colloid Interface Sci.}, \textbf{107}, 21--37.

\bibitem[Castillo \& Rosner(1988)]{CR88}
\textsc{Castillo, J. L. \&  Rosner, D. E.} 1988
A nonequilibrium theory of surface deposition from particle-laden, 
  dilute condensible vapour-containing laminar boundary layers.
\emph{Int.~J.~Multiphase Flow}, \textbf{14}, 99--120.

\bibitem[Castillo \& Rosner(1989)]{CR89}
\textsc{Castillo, J. L. \&  Rosner, D. E.} 1989
Theory of surface deposition from a unary dilute 
vapour-containing steam, allowing for condensation within the laminar boundary layer.
\emph{Chem. Eng. Sci. } \textbf{44}, 925--937.

%\bibitem[Castillo \& Rosner(1989b)]{CR89b}
%\textsc{Castillo, J. L. \&  Rosner, D. E.} 1989
%Equilibrium theory of surface deposition from a particle-laden dilute, saturated vapor 
%containing laminar boundary layers. \emph{Chem. Eng. Sci. } \textbf{44}, 939--956.

\bibitem[Davis(1983)]{davis}
\textsc{Davis, E.~J.} 1983
Transport Phenomena with Single Aerosol Particles.
\emph{Aerosol Sci. Technol.} \textbf{2}, 121--144.
  
\bibitem[Delale \& Crighton(1998)]{del98}
\textsc{Delale, C.~F. \& Crighton, D.~G. } 1998
Prandtl-Meyer flows with homogeneous condensation. Part 1. Subcritical flows. 
\emph{J. Fluid Mech.} \textbf{359}, 23--47.

\bibitem[Filippov(2003)]{fil03}
\textsc{Filippov, A.~V.} 2003
Simultaneous particle and vapour deposition in a laminar boundary layer.
\emph{J.~Colloid Interface Sci.} \textbf{257}, 2--12.

\bibitem[Garc\'{\i}a Ybarra \& Castillo(1997)]{pedro}
\textsc{Garc\'{\i}a Ybarra, P.~L. \& Castillo, J.~L. } 1997
Mass transfer dominated by thermal diffusion in laminar boundary layers. 
\emph{J. Fluid Mech.} \textbf{336}, 379--409.

\bibitem[G\"okoglu \& Rosner(1986)]{GR86}
\textsc{ G\"okoglu, S.~A. \& Rosner, D.~E.} 1986
Thermophoretically augmented mass transfer rates to solid walls across 
laminar boundary layers. \emph{AIAA~J.} \textbf{24}, 172--179.

\bibitem[Luo et al(2007)]{luo07}
\textsc{Luo, X.~S., Lamanna, G., Holten, A. P. C. \& van Dongen, M. E. H.} 2007
Effects of homogeneous condensation in compressible flows: Ludwieg-tube experiments and 
simulations. \emph{J. Fluid Mech.} \textbf{572}, 339--366 .

\bibitem[Paoli, Helie \& Poinsot(2004)]{pao04}
\textsc{Paoli, R., Helie, J. \& Poinsot, T.} 2004
Contrail formation in aircraft wakes. \emph{J. Fluid Mech.} \textbf{502}, 361--373.

\bibitem[Peeters et al(2001)]{peeters}  
\textsc{Peeters, P., Luijten, C.C.M. \& van Dongen, M.E.H.} 2001
Transport Phenomena with Single Aerosol Particles.
\emph{Int. J. Heat Mass Transfer} \textbf{44}, 181--193.

\bibitem[Pyyk\"onen \& Jokiniemi(2003)]{pyy03}
\textsc{Pyyk\"onen, J. \& Jokiniemi, J.} 2003
Modelling alkali chloride superheater deposition and its implications.
\emph{Fuel Processing Technol.} \textbf{80}, 225--262.

\bibitem[Rosner(2000)]{rosner}
\textsc{Rosner, D. E.} 2000
\emph{Transport processes in chemically reacting flow systems}. Dover.

\bibitem[Schlichting \& Gersten(2000)]{Hiemenz}
\textsc{Schlichting, H. \& Gersten, K.} 2000
\emph{Boundary Layer Theory}, 8th Edn. Springer.

\bibitem[Tandon \& Murnagh(2005)]{tandon}
\textsc{Tandon, P. \& Murtagh, M.} 2005
ParticleÐvapour interaction in deposition systems: 
influence on deposit morphology.
\emph{Chem. Eng. Sci.} \textbf{60}, 1685--1699.

\bibitem[Zheng(2002)]{zheng}
\textsc{Zheng, F.} 2002
Thermophoresis of spherical and non-spherical particles: a review of 
theories and experiments.
\emph{Adv. Colloid Interface Sci.}
\textbf{97}, 253--276.
\end{thebibliography}
\end{document}